\documentclass{aa} 
\usepackage{float} 
\usepackage{graphicx}
\graphicspath{{./Figures/}}
\usepackage{txfonts}
\usepackage{amsmath, bm}	
\usepackage{amssymb}
\usepackage{nccmath}
\usepackage{subcaption}
\usepackage{listings}
\usepackage{natbib}
\bibpunct{(}{)}{;}{a}{}{,}
\usepackage{hyperref}
\hypersetup{
 colorlinks=true,
 citecolor=blue,
 linkcolor=blue,
 urlcolor=blue,
 filecolor=blue,
 pdftitle={Anisotropy from eccentric binaries},
 }
\usepackage{lineno}
\usepackage{adjustbox}
\usepackage{multirow}
\usepackage{colortbl}
\begin{document} 

 \title{Dissecting the nanohertz gravitational wave sky: Frequency-correlated anisotropy induced by eccentric supermassive black hole binaries.}
\titlerunning{GWB anisotropies from eccentric SMBHBs}
\authorrunning{Moreschi et al.}

\author{B. E. Moreschi,\inst{1,2,3}\fnmsep\thanks{e-mail: \texttt{b.moreschi1@campus.unimib.it}} \and
S. Valtolina,\inst{4,5} \and
A. Sesana,\inst{1} \and
G. Shaifullah,\inst{1,2,3} \and
M. Falxa,\inst{1} \and
L. Speri,\inst{6} \and
D. Izquierdo-Villalba,\inst{1,2} \and
A. Chalumeau\inst{7}}

\institute{Dipartimento di Fisica “G. Occhialini”, Università degli Studi di Milano-Bicocca, Piazza della Scienza 3, I-20126 Milano, Italy \and
INFN, Sezione di Milano-Bicocca, Piazza della Scienza 3, I-20126 Milano, Italy \and
INAF – Osservatorio Astronomico di Cagliari, Via della Scienza 5, I-09047 Selargius (CA), Italy \and
Max Planck Institute for Gravitational Physics (Albert Einstein Institute), Callinstrasse 38, D-30167 Hannover, Germany \and
Leibniz Universit\"at Hannover, Callinstrasse 38, D-30167 Hannover, Germany \and
European Space Agency (ESA), European Space Research and Technology Centre (ESTEC), Keplerlaan 1, 2201 AZ Noordwijk, the Netherlands
\and
ASTRON, Netherlands Institute for Radio Astronomy, Oude Hoogeveensedijk 4, 7991 PD, Dwingeloo, The Netherlands\\
}

\date{Received 17 June 2025 / Accepted 7 April 2026}

\abstract{Revealing the nature of the nanohertz gravitational wave (GW) signal recently reported by pulsar timing array (PTA) collaborations around the world is the next goal of low-frequency GW astronomy. The signal likely originates from the incoherent superposition of GWs emitted by a cosmological population of supermassive black hole binaries (SMBHBs). These binaries can be highly eccentric and/or strongly coupled to their nuclear environment, resulting in an attenuation of the overall GW signal at low frequencies. In this paper, we propose to use the correlation properties of the distributed GW power in the sky across the frequency spectrum as a smoking gun for eccentric SMBHBs, thus allowing us to break the spectral degeneracy between eccentricity and environmental effect. The simple underlying idea is that, contrary to circular binaries, eccentric ones emit a broadband spectrum thus resulting in similar sky maps at different frequencies. We first demonstrate the applicability of this simple concept on sky maps constructed directly from the theoretical sky distribution of the gravitational wave background (GWB) power induced by realistic populations of SMBHBs. We then demonstrate the viability of this analysis on simulated square kilometer array (SKA) like PTA data. By statistically comparing sky maps reconstructed from hundreds of injected circular and highly eccentric SMBHB populations, we find that eccentricity can be detected at $3\sigma$ in more than $50\%$ of cases. }

\keywords{gravitational waves - 
       pulsars: general - 
       quasars: supermassive black holes -
       methods: data analysis -
       techniques: miscellaneous.
        }

\maketitle
\nolinenumbers
\section{Introduction \label{sec:intro}}

The increasingly strong evidence for a gravitational wave background (GWB) in pulsar timing array (PTA) experiments has intrigued the scientific community worldwide. Multiple regional PTA collaborations have reported different levels of confidence in the evidence of a common low-frequency signal in their latest data releases \citep{EPTA_DR2, CPTA_gwb, NANOGrav_15yr, PPTA_gwb, meerkat}, which show cross-pulsar correlation properties compatible with those expected for a stochastic GWB \citep{HD}. Adding more pulsars with longer observation time spans will soon bring PTA experiments to the sensitivity required for robust astrophysical inference about the nature of the low-frequency gravitational wave (GW) sources.

In short, PTA experiments study the nanohertz (nHz) frequency range, where the dominant signal is expected to be a stochastic GWB. This signal affects the times of arrival (ToAs) of pulsar observations as a low-frequency noise that is spatially correlated between different pulsars, as predicted by the Hellings and Downs (HD) correlation function \citep{HD}. The GWB can be generated by superposition of periodic GWs emitted by a cosmic population of supermassive black hole binaries \citep[SMBHBs;][]{Rajagopal_1995, jb2003, wl2003, high_freq, rosado},  although it could arise from a variety of cosmological phenomena such as cosmic strings networks \citep{2000PhRvL..85.3761D}; a non-standard inflationary scenario breaking the slow-roll consistency relations \citep{bartolo, boyle, sorbo}; cosmic domain walls \citep{hiramatsu}; primordial curvature perturbations \citep{tomita, matarrese}; or even quantum chromodynamics (QCD) phase transitions \citep{kosowsky, hindmarsh}. In this article, we work under the assumption that the sought GWB signal is produced by a population of SMBHBs and study how the discrete nature of the GW spectrum produced by this population leads to measurable anisotropies, provided a sufficiently sensitive PTA.

Standard search pipelines used in PTA analysis \citep{Ellis_2019} model the GWB as a stationary Gaussian, isotropic  process characterized by a power spectral density (PSD) well modeled by a power law. This description of the GW spectrum is well justified for an idealized population of circular binaries uniformly distributed over the sky \citep{high_freq}.

However, none of these statistical properties is expected to accurately describe the GWB signal produced by a realistic SMBHB population. Finite number statistics, environmental effects such as gas and stellar hardening, and eccentricity are expected to produce significant deviations of the spectra from the smooth power law approximation \citep{high_freq, ravi, sampsoncornish, B_2023, Valtolina_2024}. Moreover, the discreteness of the SMBHB population and/or strong individual sources may produce extra power at specific frequencies, also resulting in highly anisotropic and non-Gaussian signals \citep{sesanavecchiovolonteri, ravi2012, kelley2017, Ferranti_2025, b_2025}. In addition, the eccentricity of SMBHBs might break the assumption of stationarity \citep{Sesana_2013, falxa2024eccentricbinariesnonstationarygravitational}. Finally, resolving individual sources remains a high priority target for PTAs \citep{ng15yrCGW,EPTA_5, meerkat}, with some suggestions that the observed nHz GWB sky already has some degree of anisotropy \citep{Grunthal_2024}. 

One of the key elements to distinguish between a cosmological GWB and an astrophysical background is the expected presence of anisotropic features. Although extremely faint anisotropies ($\lesssim 10^{-20}$ $h$, where $h$ is the dimensionless strain) may arise from intrinsic features \citep{Bartolo_2022} or kinematic Doppler-like effects \citep{Jim_nez_Cruz_2024} in the GWB from cosmological sources, these signals are expected to be largely isotropic \citep{Caprini_2018}. In contrast, an astrophysical GWB is expected to exhibit greater anisotropy, driven by the discrete nature of the SMBHB population combined with galaxy clustering effects, which produce more pronounced overdensities of GW power. Cosmic microwave background (CMB) analysis techniques have been adapted to the PTA scenario to model these anisotropic features \citep{Olmez_2012, Mingarelli_2013, Gair_2014, rosado}. The idea is to model the GW power as a function of the sky location; this can be done using different bases, where the most common ones are the pixel basis, the spherical harmonics basis, and the square-root spherical harmonics basis \citep{Mingarelli_2017, rut}. Applications of these techniques to real PTA datasets have been carried out by PTA collaborations without assessing any definitive result in favor of an anisotropic GWB model over an isotropic model \citep{Taylor_2013, taylor, NG15aniso, Grunthal_2024}.

In addition to discriminating between cosmological and astrophysical origins, the measured degree of anisotropy can also be used to constrain the properties of the emitting GW sources. While \cite{pol} and \cite{depta2024} assess the sensitivity of a PTA dataset to anisotropic features of a GWB signal, many studies have investigated the relation between different astrophysical models for galaxy cluster evolution and SMBHB formation and the results of PTA anisotropy studies, using both analytical predictions \citep{satopolito2023, Allen_2024, Grimm_2025, cusin2025} and numerical simulations \citep{Gardiner_2023, lemke2024, Sah_2024, semenzato2024}.

In this paper, we explore the possibility of using anisotropy to discriminate between SMBHB populations affected by strong environmental effects and SMBHB populations featuring binaries with significant eccentricity. Both eccentricity and environmental effects are expected to absorb some of the GW power at the lowest frequencies, which makes the PSD of the GWB deviate significantly from the power law model \citep{Sesana_2013, ravi}. Thus, from the shape of the PSD only it can be extremely hard (if at all possible) to distinguish between the two scenarios. Here, we introduce a possible strategy to discriminate between them on the basis of the different imprints they leave on the GWB anisotropy as a function of frequency. 
The basic underlying idea is that circular GW sources emit monochromatically with a frequency equal to twice their orbital frequency. Therefore, sky maps reconstructed at different frequencies will feature uncorrelated anisotropies because different binaries are dominant in each sky map. Conversely, eccentric binaries emit GWs over many harmonics, which is expected to induce some level of correlation between sky maps of GW power at different frequencies. In other words, the same source will be detectable in more than one sky map.
To implement and test this idea, from the spherical harmonic decomposition of the GWB signal, we reconstruct the sky map of GW power at different frequencies \citep{Mingarelli_2013, Gair_2014}, and then compute their correlation.

The idea of cross-correlating sky maps has recently been explored in \cite{semenzato2024}, \cite{cusin2025}, and \cite{sah_2} to quantify the relation between the GWB sky power distribution and galaxy large-scale structures. Our approach is different in the sense that we are decomposing the nHz GW signal into several sky maps at different frequencies and computing their cross-correlation. In practice, the GW signal is cross-correlated with itself rather than with the distribution of galaxies and large-scale structures in the sky. As we were completing this work, \cite{Sah:2025dmv} presented an independent parallel investigation that touches on a few of the points examined here. In this article, in addition to showing the theoretical idea, we go a step further by testing our methodology using state-of-the-art SMBHB catalog realizations injected into realistic PTA simulated datasets.

The paper is structured as follows. In Section~\ref{methods}, we describe the tools upon which our analysis is built, including techniques for decomposition and cross-correlation of sky maps and models used to generate the SMBHB populations.
In Section \ref{theoretical} we present the key concept of cross-correlating GW power at different frequencies to detect eccentricity in the theoretical GW power distributions obtained from the theoretical SMBHB populations. 
In Section \ref{pta simulated dataset} we carry out a more realistic experiment by injecting the GW signal into an idealized PTA. We describe our PTA simulation approach, define the detection statistics involved in the analysis, and present the main results. Finally, in Section \ref{discussion} we comment on the merits and limitations of our investigation and point to future directions for the further development of this analysis technique.

\section{Analysis methods}
\label{methods}

In this section, we present the main ingredients of our analysis. These include strategies to decompose and cross-correlate different sky maps, as well as models adopted to simulate the SMBHB populations we are considering.

\subsection{Construction and correlation of GW power sky maps \label{gwb sky maps}}
A way to distinguish eccentric and quasi-circular populations is by inspecting and correlating the distribution of the GWB power across the sky at different frequencies. If we suppose that we evaluate the GW signal over an array of discrete frequencies $f_j$, the GWB PSD at each frequency is written as
\begin{equation}
\label{eq:power}
S_h (f_j) = \frac{h_c^{2}(f_j)}{12 \pi^{2} f_j^{3}},
\end{equation}
where $h_c(f_j)$ is the characteristic strain of the GW signal, which we will introduce in Section \ref{GWB computation}. At each $f_j$, the GWB PSD can be factorized as
\begin{equation}
\label{eq:power2}
S'_{h} (\hat\Omega,f_j) = S_{h}(f_j) \frac{P(\hat\Omega, f_j)}{4\pi},
\end{equation}
where $S_{h}(f_j)$ describes the spectral content of the GWB, and $P(\hat\Omega, f_j)$ is the distribution of the GWB power in the sky such that $\int d\hat\Omega P(\hat\Omega,f_j)=4\pi\,\, \forall f_j$.
It is possible to decompose the GWB power using either the spherical harmonics or the square-root spherical harmonics bases\footnote{While the choice of basis can be arbitrary, these are the two most common in the literature.}. 
With the former method, we write \citep{taylor}
\begin{equation}
\label{eq:alm}
P_{\text{spha}}(\hat{\Omega}, f_j) = \sum_{\ell=0}^{\infty} \sum_{m=-\ell}^{\ell} a_{\ell m}(f_j) Y_{\ell m}(\hat{\Omega}),
\end{equation}
where $Y_{\ell m}$ are the real valued spherical harmonics with coefficients $a_{\ell m}$.

The number of $l$-multipoles that can be resolved by a PTA is limited by the number of observed pulsars such that $l_\text{max} \lesssim \sqrt{N_{psrs}}$ \citep{romano}. Since we simulate $200$ pulsars (as described in Section \ref{pta simulated dataset}), we set $l_{\text{max}}=10$ \footnote{Setting $l_{\text{max}}=10$ gives us a sky resolution $\theta = 180\deg / l = 18\deg$. We also tested $l_{\text{max}}=8,12$ for all analyzes reported in this work, obtaining consistent results.}, and the sky map resolution $N_{\text{side}}=8$.

Since the decomposition of the spherical harmonics allows a negative GWB power, which is unphysical, we can use as an alternative method the square root spherical harmonic basis proposed in \citet{rut}, so that we can write the GWB power as

\begin{equation}
\label{eq:sqrt_harmonic}
{P}_{\text{sqrt}}(\hat{\Omega}, f_j)^{1/2} = \sum_{L=0}^{\infty} \sum_{M=-L}^{L} b_{LM}(f_j) Y_{LM} (\hat{\Omega}),
\end{equation}
where $Y_{LM}$ are the real valued spherical harmonics and $b_{LM}$ are coefficients of square-root decomposition, related to $a_{\ell m}$ as

\begin{equation}
a_{\ell m} = \sum_{LM}\sum_{L^{\prime}M^{\prime}}b_{LM}b_{L^{\prime}M^{\prime}}\beta_{\ell m}^{LM,L^{\prime}M^{\prime}},
\end{equation}
where we omit the $f_j$ index for brevity of notation, and

\begin{equation}
\beta_{\ell m}^{LM,L^{\prime}M^{\prime}} = \sqrt{\frac{(2L + 1)(2L^{\prime} + 1)}{4\pi(2\ell + 1)}} C_{LM,L^{\prime}M^{\prime}}^{\ell m} C_{L0,L^{\prime}0}^{\ell 0},
\end{equation}
and $ C_{LM,L^{\prime}M^{\prime}}^{\ell m} $ are the Clebsch-Gordan coefficients.
Thus, using square root spherical harmonics, we decompose ${P}_{\text{sqrt}}({\Omega}, f_j)^{1/2}$ instead of ${P}_{\text{spha}}(\hat{\Omega}, f_j)$ so that the GWB power is always positive.

Once the $j$-th map is decomposed into spherical harmonics, we compute the $C_{\ell}$ coefficients of its $\ell$ mode decomposition, representing the GWB power distribution across different angular scales corresponding to $\theta=180^\circ/\ell$
\begin{equation}
 C_\ell^{j-map} = \frac{1}{2\ell + 1} \sum_{m=-\ell}^{\ell} |a_{\ell m}^{j}|^2.
\end{equation}

We then compute the cross-correlation of each $C_\ell$ between maps $i$ and $j$ as
\begin{equation}
C_\ell^{ij}= \frac{1}{2\ell + 1} \sum_{m=-\ell}^{\ell} a_{\ell m}^{i} \left(a_{\ell m}^{j}\right)^*.
\end{equation}

Finally, we introduce the normalized correlation coefficient between the two maps $r_\ell^{ij} $
\begin{equation}
  r_\ell^{ij} = \frac{C_\ell^{ij}}{\sqrt{C_\ell^{i} C_\ell^{j}}},
\end{equation}
and a global weighted correlation $\Bar{r}^{ij} $ defined as
\begin{equation}
\label{eq:equation}
\bar{r}^{ij} = \frac{\sum_{\ell=0}^{\ell_{\text{max}}} (2\ell + 1) r_\ell^{ij} }{\sum_{\ell=0}^{\ell_{\text{max}}} (2\ell + 1)},
\end{equation}
thus taking into account all values of $\ell$ and $m$; $\bar{r}^{ij} $ is normalized so that it assumes values between $[-1,+1]$. Note that although the procedure to evaluate the correlation $\bar{r}^{ij} $ only requires the $a_{\ell m}$ coefficients, we also introduced in Eq.~\eqref{eq:sqrt_harmonic} the square-root basis decomposition because in the realistic experiment discussed in Section \ref{pta simulated dataset} we use this basis to construct sky maps using the frequentist approach described in \cite{pol}. We then decompose this map into standard spherical harmonics according to Eq.~\eqref{eq:alm} and proceed with the correlation computation.

Values of $\bar{r}^{ij} $ near $+1$ or $-1$ indicate a strong correlation or anti-correlation, while values near zero imply no correlation between maps. In our specific problem, the expectation is that 
$\bar{r}\simeq1$ when correlating sky maps produced by populations of eccentric SMBHBs, since eccentric bright sources emit in multiple harmonics. Conversely, we anticipate $\bar{r}\simeq0$ if we correlate sky maps of quasi-circular populations since different sky maps are dominated by distinct quasi-monochromatic\footnote{\footnotesize{In our "circular" population, binaries are in fact quasi-circular (generally eccentricity $e<0.1$) due to realistic evolution via three body scattering. Their GW emission is therefore not strictly monochromatic but it is nonetheless overwhelmingly dominated by a single GW harmonic (at twice the binary orbital frequency).}} binaries.

\subsection{SMBHB populations}
\label{populations}
In order to study the possibility of discriminating between eccentric and quasi-circular binaries subjected to environmental effects, we need to construct cosmic populations of SMBHBs that feature both. We therefore need a model for the cosmic SMBHB merger rate and for the dynamical evolution of individual binary systems.

For the cosmic SMBHB merger rate, we consider one of the observation-based models constructed in \cite{Sesana_2013,rosado}. In short, the SMBH merger rate is anchored to the parent-galaxy merger rate, assuming an SMBH-host galaxy relation from the literature. The host galaxy merger rate is instead constructed by multiplying the redshift-dependent galaxy mass function by the galaxy pair fraction and dividing it by an estimated merger timescale of the pairs. The mass function and pair fractions are directly measured from observations and taken from \citet{tom} and \citet{Bundy}, respectively. Conversely, the merger timescales are measured from simulations, and we use estimates from \citet{lotz}. Finally, the galaxy merger rate is converted to an SMBHB merger rate using the observed $M_{\rm BH}-M_\text{bulge}$ relation of \citet{kormendy}, thus obtaining the MBHB merger density $dn$ per unit of primary MBH mass $M_1$, mass ratio $q$, and cosmic source frame time $t_r$: $d^{3}n/(dt_r\,dM_1dq)$. One can then write $n=dN/dV_c$ and multiply by the comoving volume shell $dV_c/dz$ and by the time spent by a source at each rest-frame Keplerian log-frequency $dt_r/d{\rm ln}f_{K,r}$ to obtain the number of SMBHBs emitting per unit mass, mass-ratio, redshift, and rest-frame frequency: $d^{4}N/(dz\,dM_1\,dq\,d{\rm ln}f_{K,r})$.

It is in this last step that the SMBHB dynamics kick in. In fact, the binary frequency evolution $df_{K,r}/dt_r$ is driven both by the interaction with the environment and the GW backreaction, and depends on the eccentricity (which we have ignored thus far). If we assume that SMBHB hardening proceeds by scattering of background stars, we can write the evolution for $f_{K,r}$ and $e$ as \citep{Truant_2025}
\begin{equation}\label{eq:frequency_Evolution}
\begin{split}
\frac{df_{K,r}}{dt_r} & \,{=}\, \left( \frac{df_{K,r}}{dt_r} \right)_{*} \, {+} \, \left( \frac{df_{K,r}}{dt_r} \right)_{\rm GW} \,  \, \\ &
\,{=}\, \frac{3G^{4/3} M^{1/3} H \rho }{2(2\pi)^{2/3} \sigma}f_{K,r}^{1/3} \, {+} \, \frac{96(G \mathcal{M})^{5/3}}{5c^5} (2\pi)^{8/3} f_{K,r}^{11/3}\mathcal{F}(e),
\end{split}
\end{equation} 
\vspace{-0.5cm}
\begin{equation} \label{eq:eccentricity_Evolution}
\begin{split}
\frac{de}{dt_r} & \,{=}\, \left( \frac{de}{dt_r} \right)_{*} \, {+} \, \left( \frac{de}{dt_r} \right)_{\rm GW} \,  \, \\ & 
\,{=} \, \frac{G^{4/3}M^{1/3} \rho H K }{(2\pi)^{2/3} \sigma} f_{K,r}^{-2/3} \, {-} \, \frac{(G \mathcal{M})^{5/3}}{15c^5} (2\pi f_{K,r})^{8/3} \mathcal{G}(e),
\end{split}
\end{equation}
where $M$ and ${\cal M}$ are the SMBHB total mass and chirp mass respectively, $\sigma$ and $\rho$ are the velocity dispersion and stellar density of the host galaxy at the binary influence radius, $H$ and $K$ are dimensionless parameters calibrated against three-body scattering experiments \citep{Sesana2006}, and the subscripts $*$ and $GW$ denote the stellar hardening and GW backreaction components. Here, $\mathcal{F}(e)$ and $\mathcal{G}(e)$ take the form 
\begin{equation}
\begin{split}
\mathcal{F}(e) &\,{=}\, \frac{1+(73/24)e^2 + (37/96)e^4}{(1-e)^{7/2}}, \\ 
\mathcal{G}(e) &\,{=}\, \frac{304e + 121e^3}{(1-e^2)^{5/2}}. 
\end{split}
\end{equation}
To complete the dynamical evolution model, we need to specify $\rho$, $\sigma$, and an initial eccentricity $e_0$ at a given initial orbital frequency $f_0$. For each binary mass, the fiducial values of $\rho$, $\sigma$, and $f_0$ are computed following the procedure described in \cite{2010ApJ...719..851S}, and we assume that all binaries share the same eccentricity at formation $e_0(f_0)$. Note that each binary evolves from $f_0$, $e_0$ via the coupled differential Eqs.~\eqref{eq:frequency_Evolution} and \eqref{eq:eccentricity_Evolution}. Therefore, any binary observed at a frequency $f_{K,r}$ has an eccentricity $e(e_0, f_0, f)$ that is uniquely set by this evolution. We can therefore characterize the final distribution of observed SMBHBs that emit GWs in the nHz band as: $d^{5}N/(dz\,dM_1\,dq\,d{\rm ln}f_{K,r}\,de)$. 

\section{Expected GWB anisotropy signature \label{theoretical}}
In this section, we infer the astrophysical properties of different SMBHB populations, investigating the expected anisotropic GWB signatures. For this, we computed the expected weighted correlation $\bar{r}^{ij}$, presented in Section~\ref{gwb sky maps}, given the theoretical $h_{c}^{2}$ distribution across the sky generated by an SMBHB population. 

\subsection{Theoretical computation of the GWB strain \label{GWB computation}}

Given the differential population of SMBHBs described in the previous section, the characteristic strain $h_c(f)$ of the overall emitted GW signal can be written as
\begin{equation} \label{eq:SGWB_Population}
  \begin{aligned}
h_c^2(f)=\int_0^{\infty} {d} z \int_0^{\infty} {d} M_1 \int_0^1 {~d} q \int_0^1 {~d} e \frac{{d}^5 N}{{~d} z {~d} M_1 {~d} q {~d} \ln f_{{k}, r} {~d}e} \times \\
\left. h^2(f_{K,r}) \sum_{n=1}^{+\infty} \frac{g\left(n, e\right)}{(n / 2)^2}\right|_{f_{{K}, r}=f(1+z) / n},
\end{aligned}
\end{equation}
where 
\begin{equation}
h(f_{K,r})=\sqrt{\frac{32}{5}}\frac{(G\mathcal{M})^{5/3}}{c^4 D} (2\pi f_{{k}, r})^{4/3}, 
\end{equation}
and $D$ is the comoving distance of the source. In practice, each source contributes through an infinite series of $n$ harmonics of the orbital frequency $f_{{K}, r}$, adjusted by $1+z$ to account for cosmological redshift and thus observed at frequency $f=nf_{{K}, r}/(1+z)$. Each harmonic is weighted by a function that depends on $e$

\begin{equation}
\begin{split}
\label{eq:gne}
g(n,e) &= \frac{n^{4}}{32} \bigg[ \bigg( J_{n-2}(ne)-2eJ_{n-1}(ne)+\frac{2}{n}J_n(ne) \\
&\quad +2eJ_{n+1}(ne)-J_{n+2}(ne) \bigg)^{2} \\
&\quad +(1-e^{2})\bigg(J_{n-2}(ne)-2J_n(ne)+J_{n+2}(ne)  \bigg)^{2} \\
&\quad +\frac{4}{3n^{2}}J^{2}_n(ne) \bigg],
\end{split}
\end{equation}
where $J_n$ is the $n$-th order Bessel function of the first kind \citep{1963PhRv..131..435P}.

\begin{table}
\caption{Astrophysical properties of the simulated populations.}
\label{tab:ecc}
\centering
\begin{tabular}{lcc}
\hline
\hline
Settings & Pop 1& Pop 2 \\ 
\hline
Eccentricity at formation $e_0$     & 0.01          & 0.9      \\ 
Density of the environment        &  $\rho$ $\times$ 10  & $\rho$ / 10  \\ 
\hline
\end{tabular}
\tablefoot{$\rho$ is the fiducial stellar density at influence radius as estimated in \cite{2010ApJ...719..851S}.}
\end{table}

\begin{figure}
  \centering
  \includegraphics[width=\linewidth]{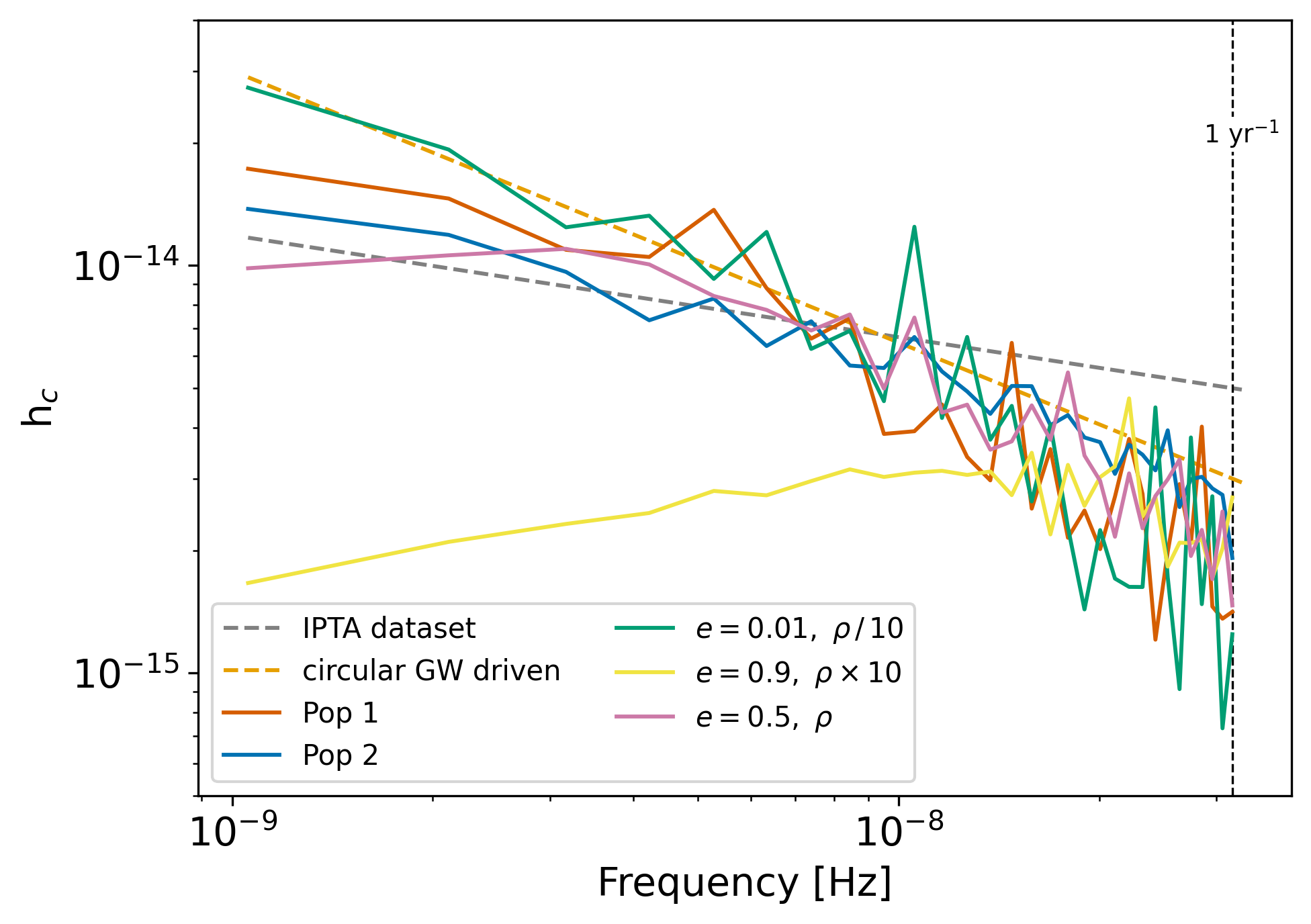}
\caption{Characteristic strain as a function of the frequency for different SMBHB populations. The dashed orange line is the GWB signal produced by a circular, GW-driven population, with $f^{-2/3}$. The dashed dark gray line is the mean GWB power law recovered using a factorized likelihood analysis of three PTA collaboration data \protect\citep{IPTA2023_Baker}. The parameters of eccentricity $e_0$ and fiducial stellar density $\rho$ at influence radius for Pop 1 and 2 are reported in Table~\ref{tab:ecc}; alternative models featuring variations of these parameters are also shown.
}
  \label{fig:fig1}
\end{figure}

\begin{figure*}
  \centering

  \begin{minipage}{0.49\linewidth}
    \centering
    \includegraphics[width=\linewidth]{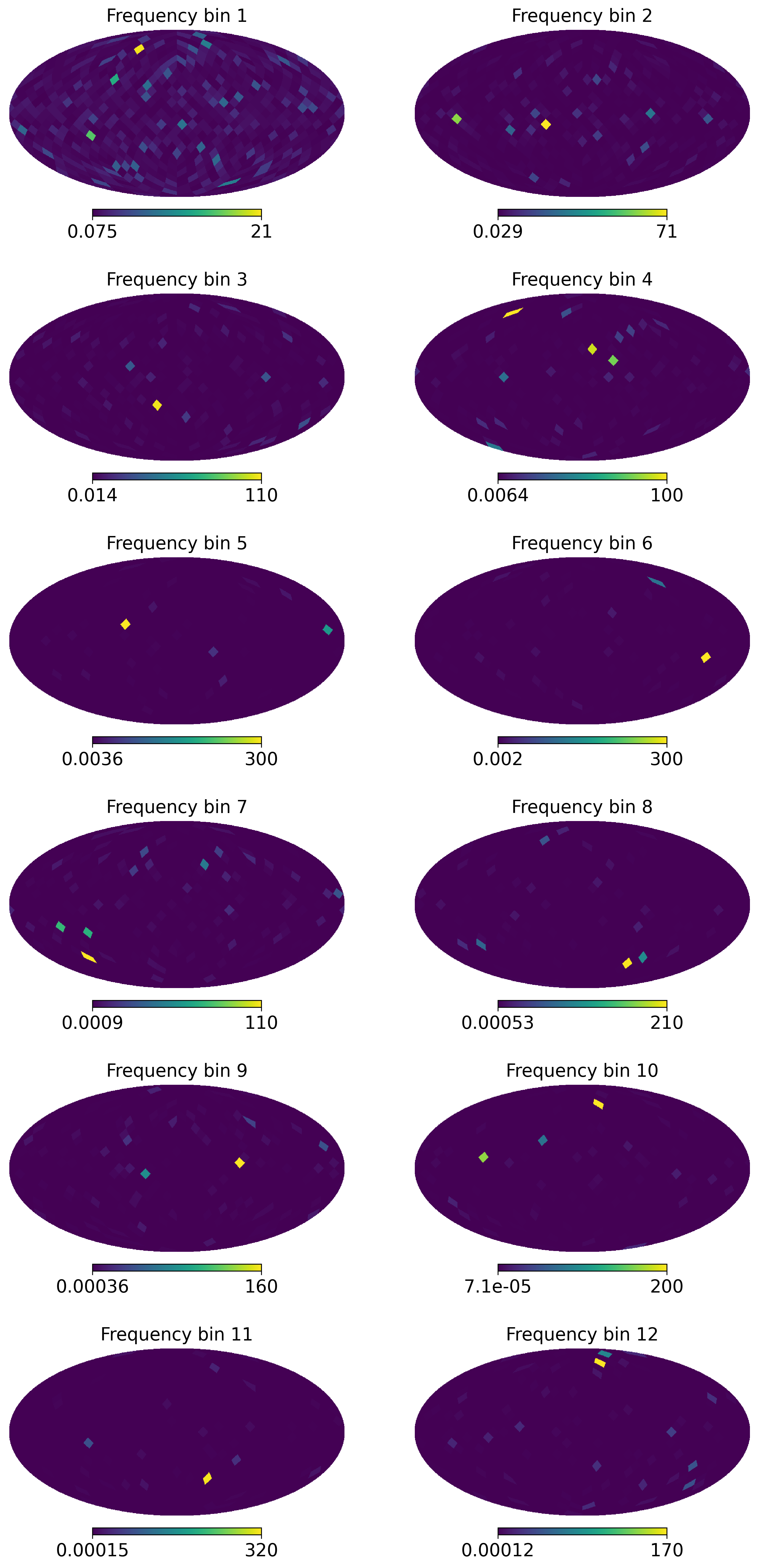}
  \end{minipage}
  \begin{minipage}[c]{0.01\linewidth}
    \centering
    \raisebox{0\height}{\rule{0.2pt}{0.7\textheight}} 
  \end{minipage}
  \begin{minipage}{0.49\linewidth}
    \centering
    \includegraphics[width=\linewidth]{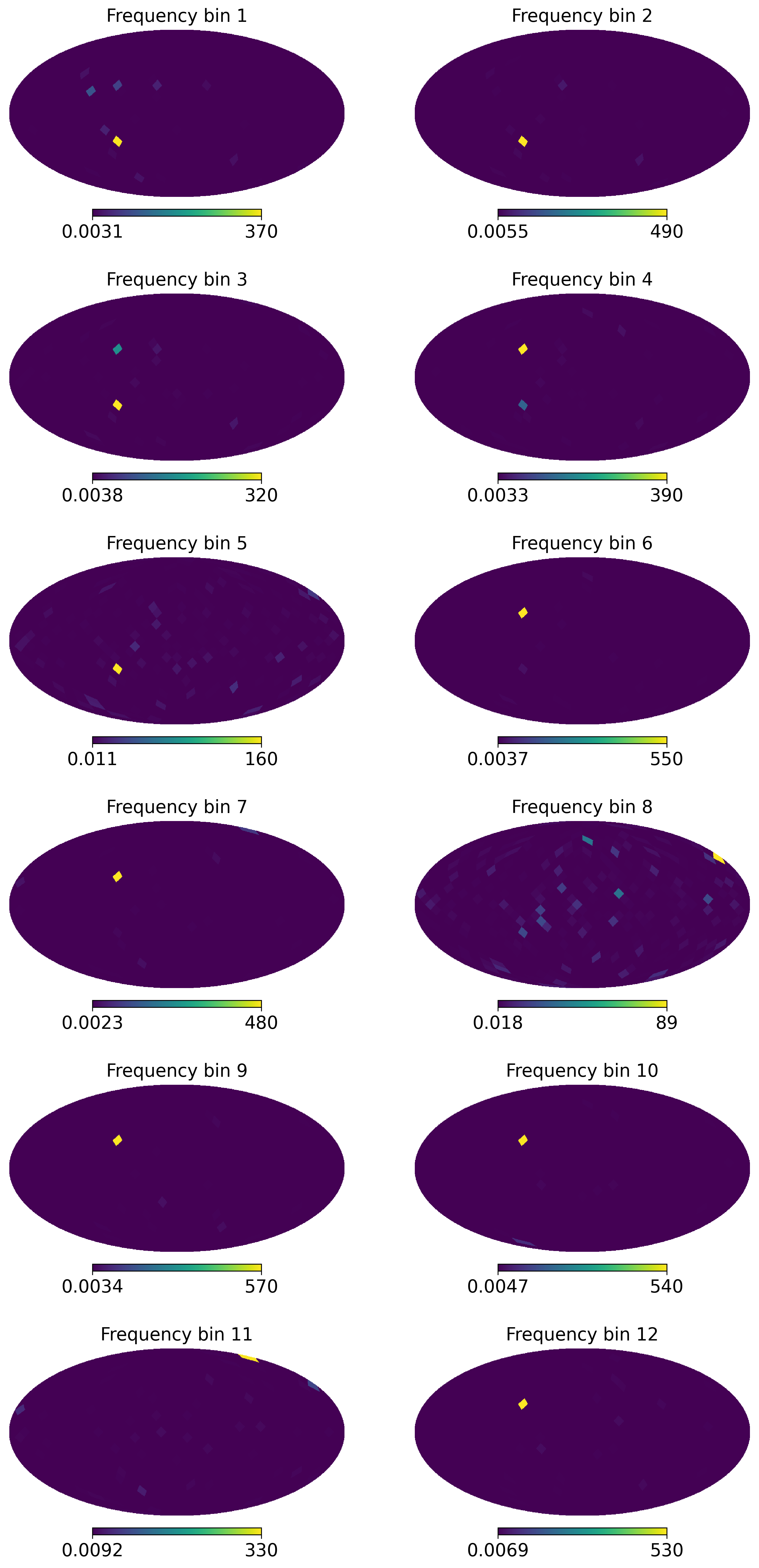}
  \end{minipage}

  \caption{
Sky maps of the GWB power distribution $P(\hat\Omega_k, f_j)$ in the first 12 frequency bins of Pop 1 (left plot) and Pop 2 (right plot). We notice that maps are often dominated by a single bright source, visible as a yellow pixel. For a quasi-circular population, different maps show bright sources located at different sky locations. On the other hand, when considering an eccentric population, the same bright source appears at the same sky location across multiple maps.
  \label{fig:sky_maps}}
\end{figure*}

Eq.~\eqref{eq:SGWB_Population} is the continuum limit solution of an inherently discrete problem. In fact, the GWB is made up of a discrete population of SMBHB and PTAs observed for a finite time $T$ and hence has a resolution limit $\Delta{f}=1/T$. This means that the GWB spectrum is evaluated in discrete frequency bins $\Delta{f_j}=[j/T, (j+1)/T]$, centered at $f_j=(2j+1)/(2T)$ for $j=1,...,N_{\rm b}$, which being $N_{\rm b}$ the number of considered frequency bins. Throughout this paper, we assume $T=30$ years. Therefore, in a real-life experiment, Eq.~\eqref{eq:SGWB_Population} takes the form \citep{2010MNRAS.402.2308A}
\begin{equation}
\label{eq:hok}
h_{c}^{2} (f_{_j}) = \sum_{i=1}^{N}\sum_{n=1}^{+\infty} h^2_i(f_{K,r})\frac{g\left(n, e_i\right)}{(n / 2)^2}\frac{n f_{K,r_i}}{\Delta f (1 + z)}\Theta\left(\frac{n f_{K,r_i}}{1+z},\Delta{f_j}\right),
\end{equation}
where $\Theta=1$ if $n f_{K,r_i}/(1+z)\in \Delta{f_j}$, and $\Theta=0$ otherwise. Here, the first sum is over the $N$ SMBHBs that make up the population, each of which contributes to the signal with the  $n$ harmonics. Note that in the continuum limit, for a population of circular GW driven binaries, Eqs.~\eqref{eq:hok} and \eqref{eq:SGWB_Population} tend to the well known power law $h_c(f)=A_{\text{GWB}}(f/{\rm 1yr}^{-1})^{-2/3}$, where $A_{\text{GWB}}$ is the nominal GWB amplitude at $f={\rm 1yr}^{-1}$\citep{phinney}. The SMBHB population described in Section \ref{populations} results in $A_{\text{GWB}}\sim3\times10^{-15}$, consistent with current PTA observations.

As proof of concept, we investigate two models that share the same SMBHB merger rate but differ in dynamical evolution due to different $e_0$ and $\rho$, as detailed in Table \ref{tab:ecc}. We refer to the models as Pop 1 and Pop 2.
For each model, we draw 100 Monte Carlo realizations of the distribution $d^{5}N/(dz\,dM_1\,dq\,d{\rm ln}f_{K,r}\,de)$. 
For each sampled source, we extracted the rest frame masses, redshift, observed GW second harmonic, eccentricity, and the $h_{c}^{2}$ contributed in each frequency bin, constructing the GWB strain amplitude according to Eq.~\eqref{eq:hok}.

The result of this procedure is shown in Figure \ref{fig:fig1} for selected realizations of the two models. There, the red and blue curves correspond to Pop 1 and Pop 2, respectively. The other solid lines show backgrounds generated by populations with different initial conditions. We notice some deviations in the background slope at low frequencies with respect to a GWB generated by a population of circular, GW-driven SMBHBs (dashed orange line). A similar flat slope has been reported by the international pulsar timing array (IPTA) \citep[dashed dark gray line]{IPTA2023_Baker}. Any deviation from the power law is due to the discrete nature of the GWs emitted by these SMBHBs, which produces a structured GWB \citep{high_freq}, as well as the level of environmental effects and the eccentricity of the sources that suppress power at low frequencies, as mentioned before. In fact, SMBHBs influenced by stellar hardening evolve faster and spend less time emitting GWs at lower frequencies. On the other hand, eccentric SMBHBs emit GWs at multiple harmonics of the orbital frequency, distributing the GW power at higher frequencies.
Although the two populations have different values of $e_0$ and different strengths of environmental coupling, it is not possible to distinguish the nature of the population from their spectra alone. This is why we investigated a new method that can be applied to real PTA datasets to determine whether the measured flat slope is due to some level of eccentricity or environmental effects (or a combination of both, although we defer the latter case to future work). 

\subsection{Cross-correlation results}
\label{results teo}

To map the distribution of the GWB power in the sky, we assigned to the $i$-th SMBHB a random sky location $\hat\Omega_i$, divided the sky into pixels $\hat\Omega_k$ of equal area $\Delta\Omega$, and evaluated the discrete version $P(\hat\Omega_k, f_j)$ of the power distribution in Eq.~\eqref{eq:power2} as 
\begin{equation}
  P(\hat\Omega_k, f_j)=\frac{h_c^{2} (\hat\Omega_k,f_j)}{h_c^2(f_j)},
\end{equation}
where $h_c^{2}(\hat\Omega_k,f_j)$ is evaluated by multiplying Eq.~\eqref{eq:hok} by $\delta(\hat\Omega_i,\hat\Omega_k)$, thus counting in each pixel only those SMBHBs that fall within it.

The distribution of $P(\hat\Omega_k, f_j)$ across the sky is shown in Figure \ref{fig:sky_maps} for the first 12 frequency bins of two selected realizations of both models (Pop 1 and Pop 2). The figure 
demonstrates the difference between quasi-circular and eccentric populations when it comes to the sky distribution of the GW power as a function of frequency. We constructed these sky maps using \texttt{HEALPY} \citep{Zonca2019} and plotted the GWB power in equal-area pixels, where $N_{\text{pix}} = 12N_{\text{side}}^{2}$. We used $N_{\text{side}}$=8 due to limitations in the number of pulsars in the PTA, as discussed above.

We distributed the sources randomly on the sky, leaving the study of potential anisotropies due to galaxy clustering for future work. For the eccentric population (Pop 2, right plot) we observe the contribution of the same source, at a specific position in the sky, emitting GWs across multiple harmonics as per Eq.~\eqref{eq:gne}, and thus in different (frequency bin-wise) sky maps. For this specific realization, we clearly see one source that dominates in bins 1, 2, 3, and 5, and another source that dominates in bins 4, 6, 7, 9, 10, and 12. It is therefore possible to note the same power distribution pattern in nearby frequency bins, and we expect a certain degree of correlation in the resulting sky maps.
Conversely, quasi-circular binaries (Pop 1, left plot) subjected to environmental effects cannot produce this kind of recursive pattern, and we do not expect the GW power between different sky maps to be correlated. This indicates that correlation across sky maps can be used to break the degeneracy of the spectrum that we saw in Figure \ref{fig:fig1}.

\begin{figure}
  \centering

  \begin{subfigure}[b]{1\linewidth}
    \centering
    \includegraphics[width=\linewidth]{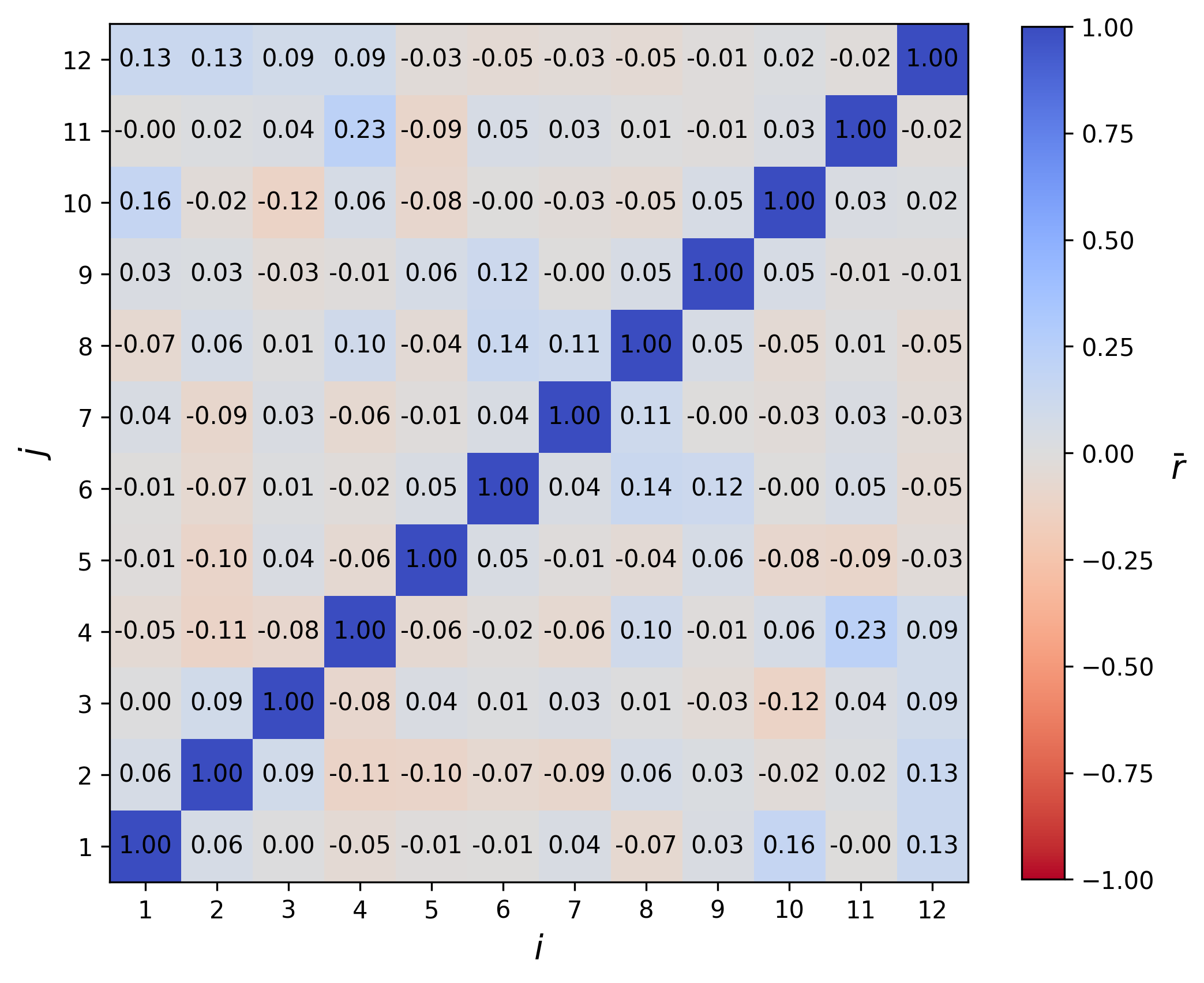}
  \end{subfigure}

  \begin{subfigure}[b]{1\linewidth}
    \centering
    \includegraphics[width=\linewidth]{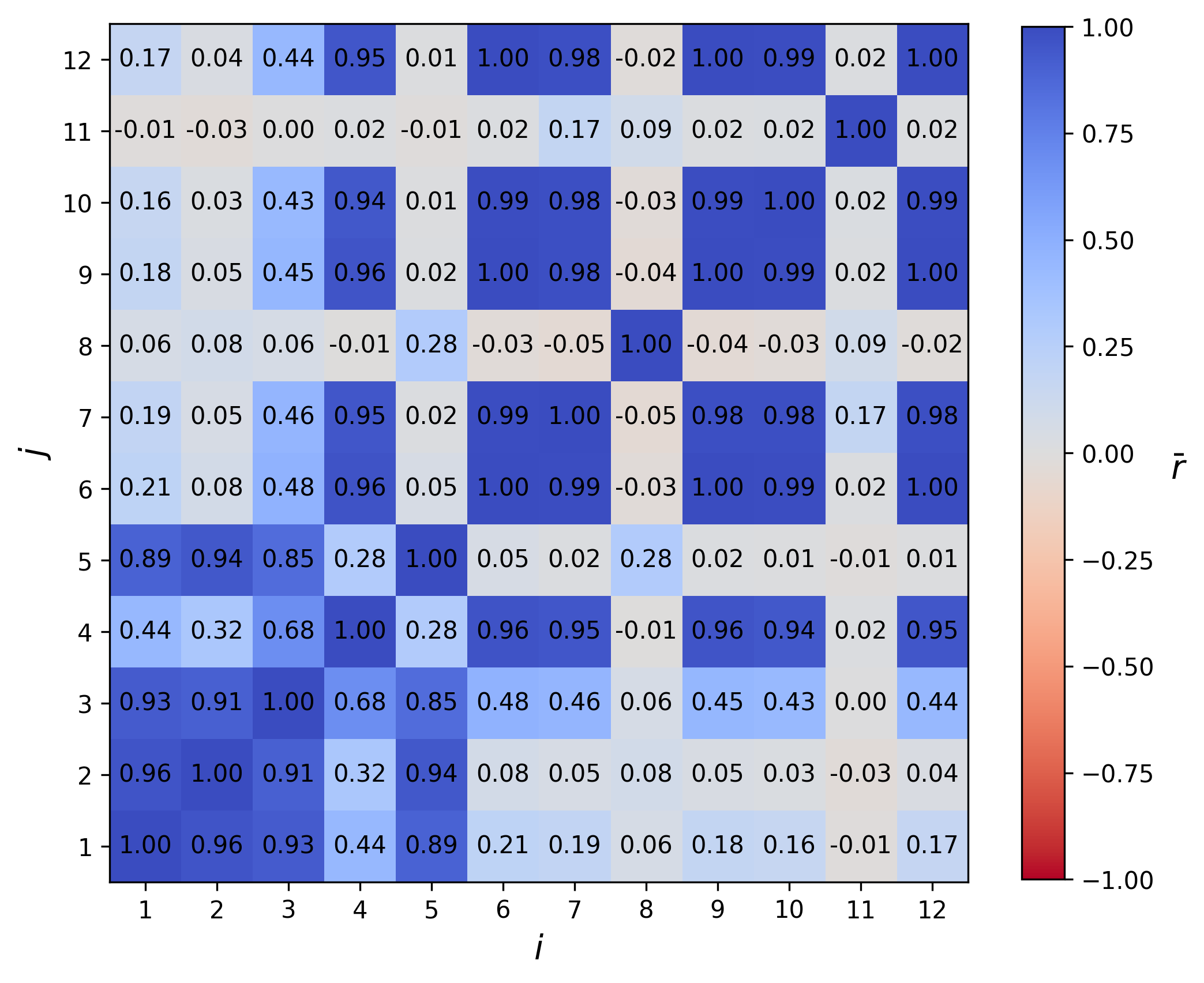}
  \end{subfigure}
\caption{
  Weighted correlation $\bar{r}^{ij}$ from a single realization of Pop 1 (top panel) and Pop 2 (bottom panel). The color scale tracks the $\bar{r}^{ij}$ values, which are also reported in each cell of the matrix. The $i$ and $j$ indices correspond to different frequency bins. \label{fig:corr_pta_theo}}
\end{figure}

To evaluate the correlation between sky maps at different frequencies $\bar{r}^{ij}$, we decomposed the power $P(\hat\Omega_k, f_j)$ into spherical harmonics according to Eq.~\eqref{eq:alm} and followed the procedure presented in Section \ref{gwb sky maps}. We evaluated $\bar{r}^{ij}$ across the 12 lowest frequency bins, thus constructing a 12$\times$12 correlation matrix.

The results for the two realizations of Pop 1 and Pop 2 displayed in Figure \ref{fig:sky_maps} are shown in Figure \ref{fig:corr_pta_theo}. Looking at the quasi-circular and strong environmental effects population (Pop 1, top panel), we can see, as expected, that the correlation is close to zero for all the off-diagonal elements of the matrix (corresponding to sky maps at different frequency bins), and equal to one on the diagonal (i.e., for an autocorrelated sky map). We obtain a similar matrix considering every quasi-circular and low environmental effect population. This happens because environmental coupling produces only a flattened spectrum at lower frequencies and not a correlation of the power when comparing different sky maps. On the other hand, if we consider Pop 2 (bottom panel), off-diagonal elements assume values higher than zero, and quite often close to unity, meaning that sky maps are correlated due to eccentric sources emitting power at different frequencies (as clearly shown in Figure \ref{fig:sky_maps}, where two bright sources dominate the GWB). Again, similar correlation maps are found for every individual realization of model Pop 2. 

\section{Recovery with simulated PTA datasets \label{pta simulated dataset}}

Although the results shown thus far are encouraging, they only concern the theoretical anisotropy of the expected GWB power. In a real PTA detection, each individual SMBHB affects the pulse ToAs of each pulsar in the array, and the reconstructed signal is a nontrivial convolution of an incoherent superposition of deterministic GWs with the pulsar response function, contaminated by noise. It is therefore of paramount importance to test the performance of our technique on realistic simulated PTA datasets.

To this end, in Section \ref{injection} we inject the residuals produced by the SMBHB populations defined in Section \ref{populations} into an ideal SKA-like PTA \cite[e.g.,][and references therein]{Truant_2025}, and we employ a frequentist technique to reconstruct from the timing residuals the sky maps of the GW power at different frequencies (Section \ref{sec:freq}). As we show, the resulting maps are much less polished compared to the theoretical ones, therefore in Section \ref{statistics} we devise statistical metrics to identify eccentricity and discuss our results in Section \ref{pta}.  

\subsection{Pulsar noise and realistic GWB injection}
\label{injection}
Our simulated array contains $200$ randomly distributed pulsars in the sky, observed with a cadence of $14$ days, over $30$ years. For each of them, we simulate ToAs using the \texttt{libstempo} python wrapper \citep{vallisneri} for the \texttt{Tempo2} pulsar timing package \citep{hem+2006,ehm+2006}. The ToA sets corresponding to the common PTA observing frequencies of 1400 and 2200~MHz are generated for each pulsar. To obtain reasonably realistic datasets that can be analyzed rapidly enough to allow several hundred realizations, we inject only white noise (WN) into the GW-perturbed ToAs, deferring a full analysis that incorporates red noise (RN) and dispersion measure (DM) variation-linked noise processes \citep[see e.g.,][]{Lentati+2013,vHLML09} to future work. The WN appears in the timing residuals with the same power at all frequencies, thus with a flat PSD, which we model as a Gaussian random process. For standard PTA analyzes, WN is composed of two terms: error-scaling factor (EFAC, $E_F$) and error in quadrature (EQUAD, $E_Q$). The former is a multiplicative parameter that scales the ToA uncertainties to account for misestimation of the ToA error-bars, while the latter adds in quadrature additional noises generated, for example, by instrumental effects or pulse profile "jitter" \citep[see][and references therein]{vs18}.
Therefore, the ToA uncertainty can be written as
\begin{equation}
\sigma = \sqrt{E_F^2 \sigma_{\text{ToA}}^2 + E_Q^2},
\end{equation}
where $\sigma_{\text{ToA}}$ is the error on the ToA. We inject WN with identical statistical properties in all pulsars, with $\sigma_{\text{ToA}}=30$ ns, EFAC = 1.0, and EQUAD = $10^{-9}$. The full setup of the simulated PTA settings is given in Table \ref{tab:tablesettings}. 

We then add to the ToAs the GW signal generated by the SMBHB populations described in Section \ref{populations}. Specifically, using a \texttt{python} and \texttt{fortran}-based pipeline, we injected in the time domain the residuals induced by each SMBHB on the ToAs of each pulsar in the PTA. In short, the code takes the initial parameters of the binary (assumed to be those at the start of observations) and evolves it forward for the observation time $T$ and backward for 20000 years using the postNewtonian equations of \citep{2004PhRvD..69h2005B}. The eccentric residuals are then injected following \cite{taylor} and \cite{Truant_2025}, both for the Earth and for the pulsar term. The latter is computed at a time $\tau_p=-(d_p/c)(1-{\bf n}\cdot{\bf n_p})$, where $d_p$ is the distance to the pulsar and ${\bf n}$,\,${\bf n_p}$ are the unit vectors pointing to the GW source and to the pulsar, respectively. In practice, we "rewind" the waveform for a time $\tau_p$ and inject the corresponding chunk into the residuals. To execute this procedure, we need to complement each SMBHB entry in the catalog with additional parameters, namely cosine of the inclination angle, polarization, and phase. These parameters are randomly sampled within their appropriate boundaries. 

Using this procedure, we create 100 mock realizations of PTA observation, for each SMBHB population model, Pop 1 and Pop 2. We note that the injected GWB in each realization has spectral properties that closely match the recovered amplitude and power law spectral index from \cite{IPTA2023_Baker}.
\begin{table}
\caption{Settings of the simulated PTA used in our analysis.}
\label{tab:tablesettings}
\centering
\begin{tabular}{cc}
\hline
\hline
Settings & SKA-like PTA \\ 
\hline
Number of pulsars          & 200      \\ 
$T$              & 30 yrs     \\
Cadence               & 14 days    \\ 
EFAC                 & 1       \\ 
EQUAD                & $10^{-9}$   \\ 
$\sigma_{\text{ToA}}$              & 30 ns     \\ 
Observation frequencies       & 1400, 2200 MHz \\ 
\hline
\end{tabular}
\end{table}

\begin{figure}
  \centering
  \includegraphics[width=\linewidth]{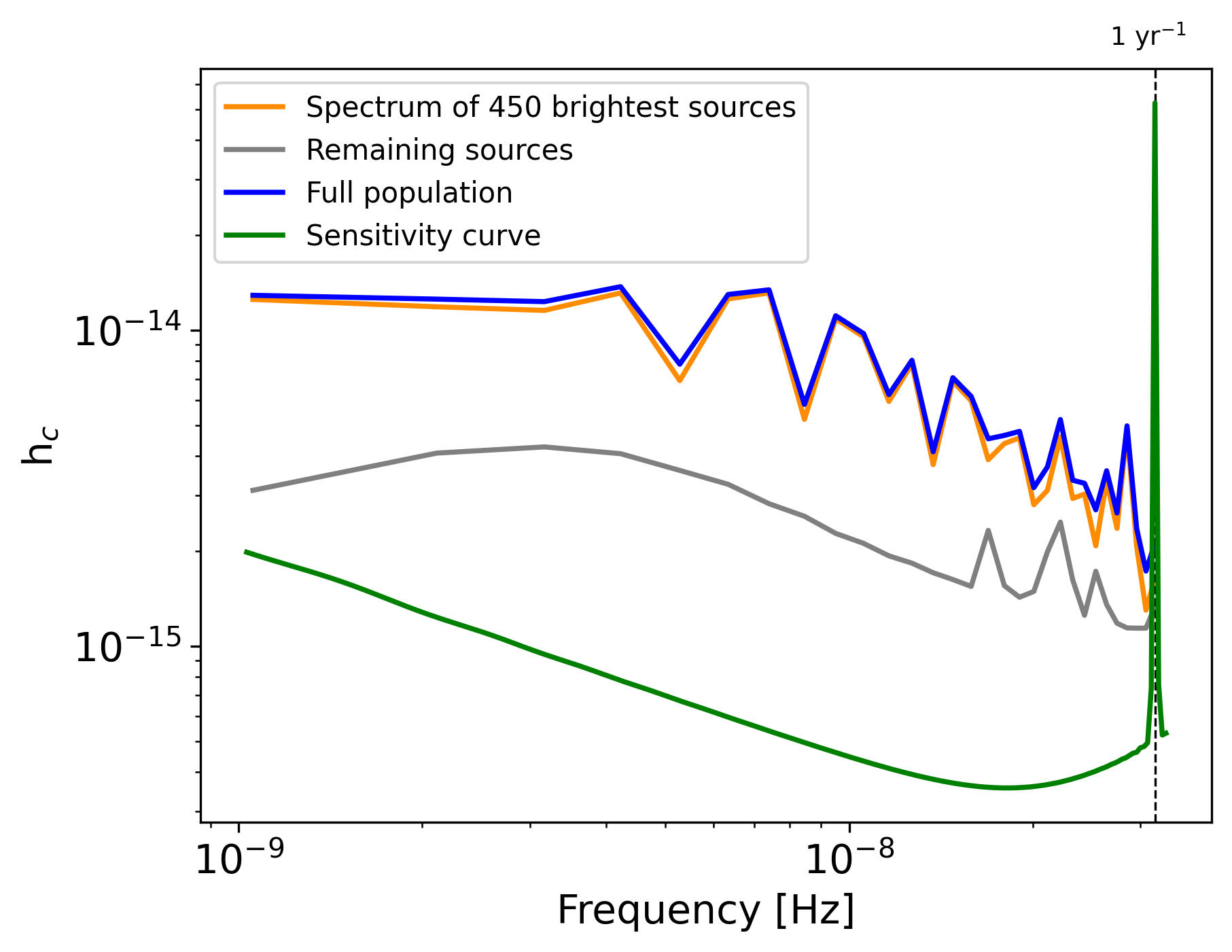}
\caption{GWB strain of a selected eccentric population injected in the PTA. The orange line is the total spectrum of the brightest sources injected individually. The gray line is the GWB spectrum of the remaining sources, which we have injected separately using the pipeline \texttt{fakepta} to ensure isotropy, and the blue line is the sum of the gray and orange lines. The green line marks the sensitivity of the simulated PTA.}
    \label{fig:spectrum}
\end{figure}

To speed up the dataset generation, especially for populations with high eccentricity, we made the following approximations. Since the GWB anisotropy is dominated by the loudest sources, while fainter sources make up a more isotropic background, we only individually inject the signals of the 30 brightest sources in each of the first 15 frequency bins (450 sources in total, instead of the $\approx 10^5$ SMBHB making up the overall population) and model the remaining sources as an isotropic GWB. The strain of the latter is computed at each frequency following Eq.~\eqref{eq:hok}, and then converted into power spectral density by applying Eq. \eqref{eq:power}. This power is then injected isotropically using a dedicated pipeline \footnote{\url{https://github.com/mfalxa/fakepta}}. Modeling the SMBHB population as an isotropic GWB plus a limited number of individually modeled bright sources per frequency bin is a robust and well-justified approximation, as also demonstrated in \cite{bence_exp}.

An example of this procedure is shown in Figure \ref{fig:spectrum}. It is clear that the GW signal produced by the subset of the loudest SMBHBs accounts for more than 90\% of the GW power, providing an excellent approximation of the properties and peculiar features of the complete population, while the remaining sources contribute to a much lower level. To ensure that this method does not bias our results, for a few selected realizations, we compared the resulting sky maps and correlation matrices to those obtained by injecting individually each SMBHB in the population, finding no significant differences.

\subsection{Frequentist GWB power reconstruction}
\label{sec:freq}

We now need to extract the GWB signal and its power distribution across the sky from the PTA data. Since current Bayesian pipelines for inference on the anisotropy coefficients $C_{\ell}$ are computationally demanding, requiring a large number of samples to reach convergence, in this analysis we use a more computationally efficient frequentist approach. In particular, we use the per-frequency optimal statistic (PFOS) tools developed by \cite{gersb}\footnote{\url{https://github.com/GersbachKa/defiant}} to model the GWB power, compute the $C_{\ell}$ values at the multipoles in the range [$0$, $ \ell_{max}$] at different GW frequencies, and consequently compute the weighted correlation $\bar{r}^{ij}$ as 
Eq.~\eqref{eq:equation}. 

While in the ``standard'' optimal statistic (OS) \citep{an}, cross-correlations between pulsars are filtered against a power law GWB spectral template, PFOS has the flexibility to accept any spectral shape. It is therefore well suited for realistic GWBs, where source discreteness, eccentricity, and environmental coupling can result in complex spectra. However, we note that the PFOS relies on the assumption that the signal is stationary, which breaks down in the case of eccentric binaries \citep{falxa2024eccentricbinariesnonstationarygravitational}. We therefore recognize that constructing sky maps with the PFOS might be suboptimal and other techniques should be explored in the future.

\label{frequentist analysis}
According to the PFOS, we can write the spatial cross-correlations between pulsars $a$ and $b$ at frequency $f_j$ and their uncertainties as

\begin{equation}
\label{eq:rho}
\rho_{ab}(f_j) = \frac{\textbf{X}_{a}^{T}\tilde{\phi}(f_j)\textbf{X}_b}{\text{tr}\Big[ \textbf{Z}_a \tilde{\phi}(f_j)\textbf{Z}_b \Phi(f_j) \Big]},
\end{equation}

\begin{equation}
\sigma_{ab,0}(f_j)^{2} = \frac{\text{tr}\Big[ \textbf{Z}_a \tilde{\phi}(f_j) \textbf{Z}_b \tilde{\phi}\Big]}{\text{tr}\Big[ \textbf{Z}_a \tilde{\phi}(f_j)\textbf{Z}_b \Phi(f_j) \Big]^{2}},
\end{equation}
where
\begin{equation}
\textbf{X}_{a} = \textbf{F}_{a}^{T} \textbf{N}_{a}^{-1} \boldsymbol{\delta t}_{a} \qquad \textbf{Z}_{a} = \textbf{F}_{a}^{T} \textbf{N}_{a}^{-1} \textbf{F}_{a}.
\end{equation}
If we collect the $N_r$ residual for the pulsar $a$ and model $J$ frequency bins (i.e. $j=1, 2, ..., J$), then $\textbf{F}_{a}$ is the $N_r\times2J$ Fourier design matrix for the pulsar $a$, $\textbf{N}_{a}$ its diagonal $N_r\times N_r$ white noise covariance matrix, and $\boldsymbol{\delta t}_{a}$ is the vector of $N_r$ timing residuals, such that $\textbf{X}_{a}$ is a vector of 2$J$ and 
$\textbf{Z}_{a}$ is a matrix of $2J\times2J$.
$\Phi(f_j)= \phi/S(f_{j})$ is the normalized Fourier domain covariance matrix, which being $\phi=\phi_{j,j'}=\delta_{j,j'}S(f_{j})\Delta{f}$. Finally, the authors introduced a new quantity $\tilde{\phi}(f_j) $ which is a frequency-selector matrix \citep[see][for full details]{pol,gersb,gersbach_2}.

If we divide the sky into $N_{pix}$ equal-area pixels $\hat\Omega_k$ the response function of each pulsar pair in the PTA to the GW power coming from a given pixel can be written as \citep{pol}
\begin{equation}
  R_{ab}(\hat\Omega_k)=\frac{3}{2}\left[{\cal F}^+_{a,j}(\hat\Omega_k){\cal F}^+_{b,j}(\hat\Omega_k)+{\cal F}^\times_{a,j}(\hat\Omega_k){\cal F}^\times_{b,j}(\hat\Omega_k)\right],
\end{equation}
such that the expectation value of the normalized cross-correlation defined by Eq.~\eqref{eq:rho} takes the form
\begin{equation}
  \langle \rho_{ab}(f_j) \rangle =\Gamma_{ab}(f_j)=\sum_k P(\hat\Omega_k, f_j)R_{ab}(\hat\Omega_k),
  \label{eq:corr_decomposition}
\end{equation}
where $P(\hat\Omega_k, f_j)$ in the GW signal power falling within the $k$-th pixel at frequency $f_j$. We can sum over pulsar pairs and write Eq.\eqref{eq:corr_decomposition} in matricial form as $\bm{\Gamma}_j=\textbf{R}\textbf{P}_j$, where $\textbf{P}_j$ is a vector of length $N_{pix}$ defining the GW power at each pixel and \textbf{R} is a $N_{cc} \times N_{pix}$ overlap response matrix where $N_{\text{cc}}=N_{\text{psr}}(N_{\text{psr}}-1)/2$ is the total number of cross-correlations.

Assuming a stationary Gaussian distribution for the cross-correlation uncertainties, we can write the likelihood function for the cross-correlation vector $\bm{\rho}_j$ as
\begin{equation}
\label{eq:likelihood}
p(\bm{\rho}_j | \textbf{P}_j) = \frac{\exp\left[-\frac{1}{2} (\bm{\rho}_j - \textbf{R} \textbf{P}_j)^{T} \bm{\Sigma}_j^{-1} (\bm{\rho}_j - \textbf{R} \textbf{P}_j)\right]}{\sqrt{\det(2\pi \bm{\Sigma}_j)}},
\end{equation}
where $\bm{\Sigma}_j$ is the noise covariance matrix $N_{\text{cc}} \times N_{\text{cc}}$ of the cross-correlations at frequency $f_j$ and $\bm{\rho}_j$ is the vector of $N_{cc}$ cross-correlations given by Eq.~\eqref{eq:rho} at each frequency bin.

In Eq.~\eqref{eq:likelihood}, we write each term $P(\hat\Omega_k, f_j)$ in the power matrix $\textbf{P}_j$ as a decomposition in square root spherical harmonics according to Eq.~\eqref{eq:sqrt_harmonic}. Therefore, by maximizing the likelihood, we get an estimate of the $b_{LM}(f_j)$ coefficients from the PTA dataset, obtaining the GWB power ${P}_{\text{sqrt}}(\hat{\Omega}, f_j)^{1/2}$. Finally, we use this decomposition to build sky maps\footnote{See notes in \url{https://github.com/NihanPol/MAPS}} that we cross-correlated according to Section \ref{gwb sky maps} to construct the correlation matrix ${\bar{r}^{ij}}$.
Following \cite{pol}, we compute the maximum-likelihood solution numerically using the LMFIT package \citep{newville}. 

Eq.~\eqref{eq:likelihood} can also be evaluated by forcing a constant $P(\hat\Omega_k, f_j)=P(f_j)_{\rm iso}$ across the sky, i.e., assuming that the signal is isotropic. We can then adopt as a detection statistic for anisotropy at each frequency the S/N defined as

\begin{equation}
\label{eq:snr}
\text{S/N}_j = \sqrt{2 \text{ln} \left[ \frac{p(\bm{\rho}_j | \textbf{P}_j)}{p(\bm{\rho}_j | \textbf{P}_{j,{\rm iso}})}\right ]},
\end{equation}
where we compare for each frequency bin the maximum likelihood of the recovered anisotropic sky map $\textbf{P}_{j}$ with an isotropic model $\textbf{P}_{j,{\rm iso}}$, considered as the null hypothesis. We find that in all our populations this S/N is above ten in the lowest frequency bin.

\begin{figure}
  \centering

  \begin{subfigure}[b]{1\linewidth}
    \centering
    \includegraphics[width=\linewidth]{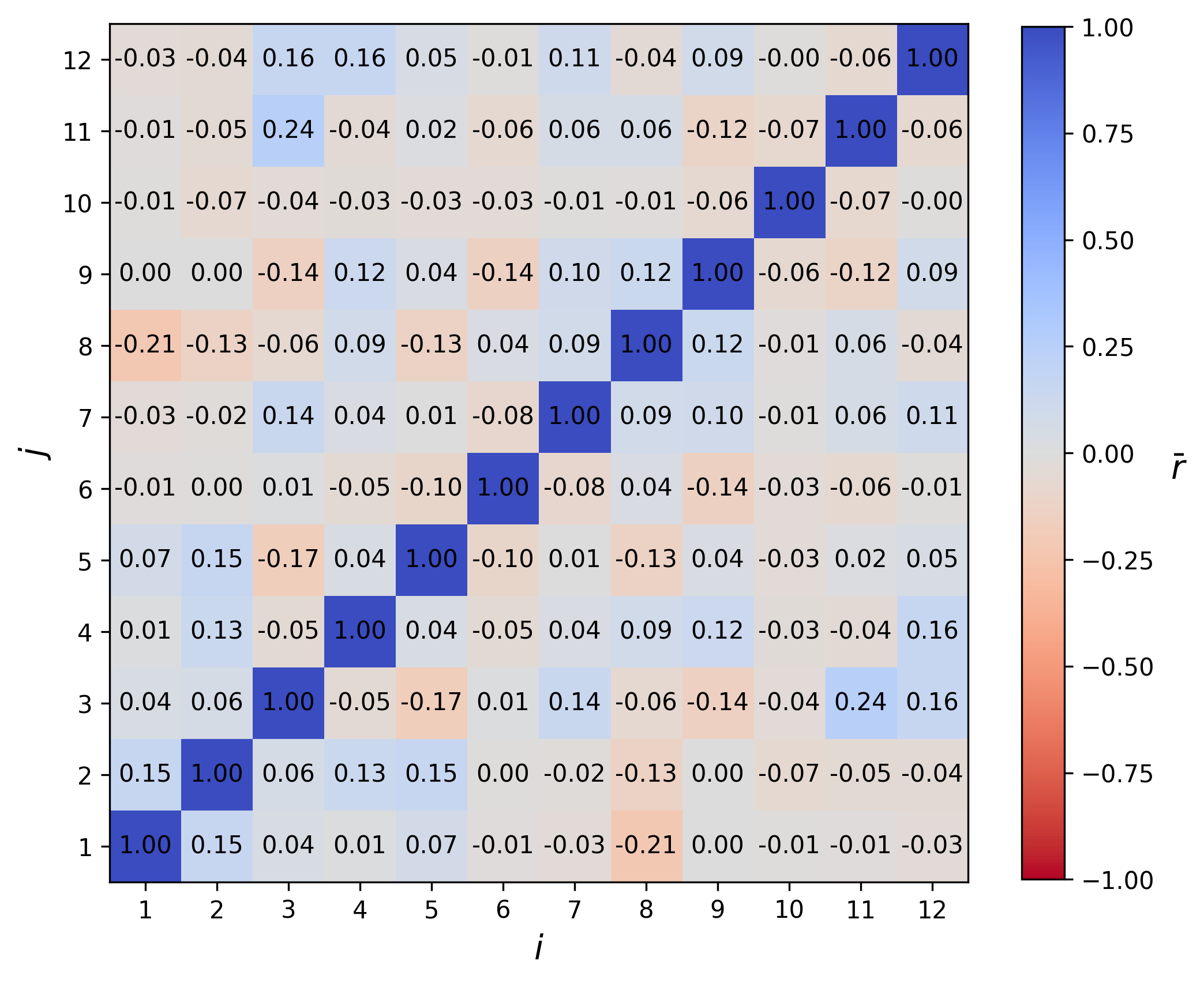}
  \end{subfigure}

  \begin{subfigure}[b]{1\linewidth}
    \centering
    \includegraphics[width=\linewidth]{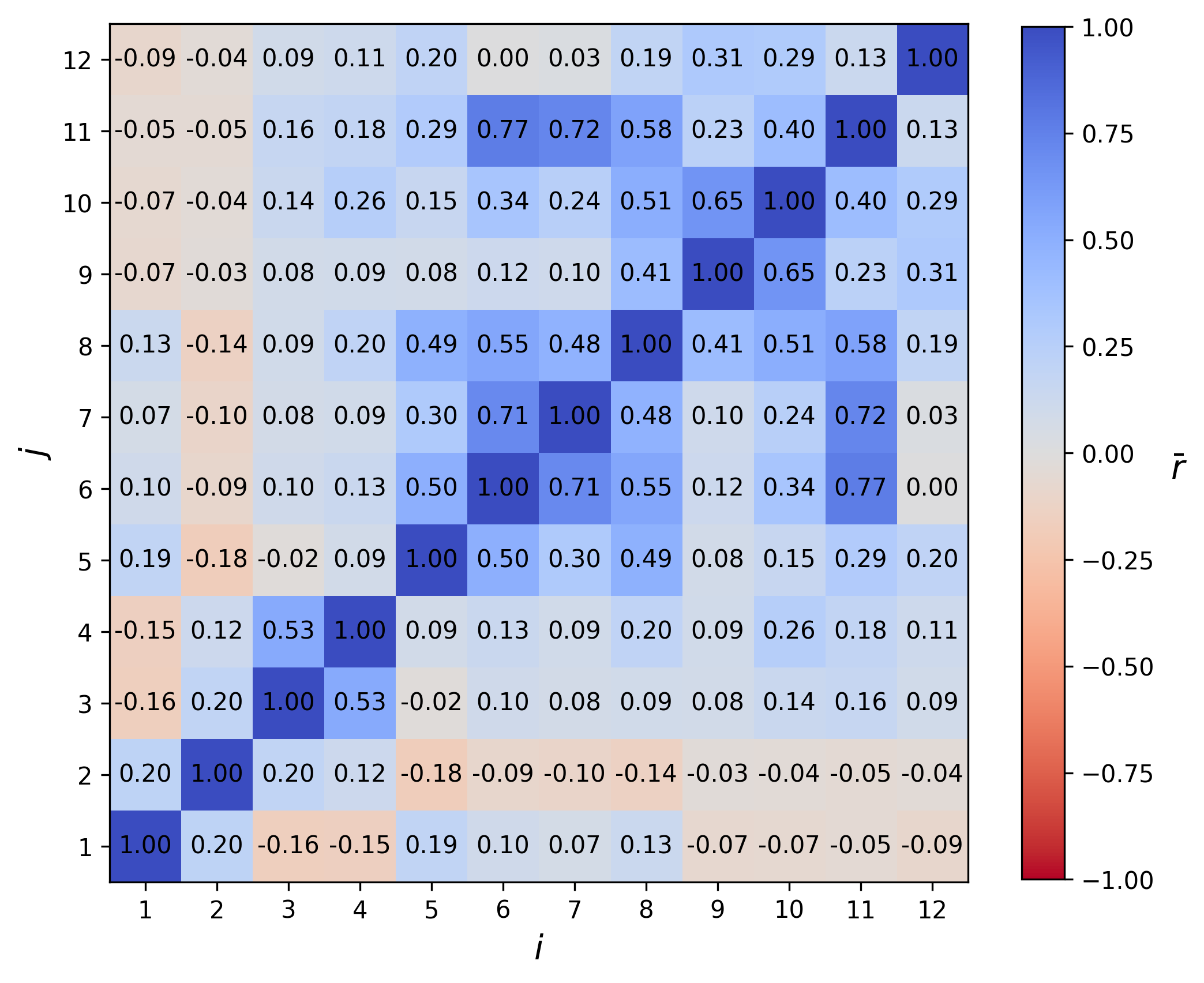}
  \end{subfigure}

\caption{
    Weighted correlation $\bar{r}^{ij}$ from a single realization of Pop 1 (top panel) and Pop 2 (bottom panel) computed from the simulated PTA dataset. The color scale tracks the $\bar{r}^{ij}$ values, which are also reported in each cell of the matrix. The $i$ and $j$ indices correspond to different frequencies $i/T$ and $j/T$.}

  \label{fig:corr_pta}
\end{figure}

\begin{figure}
  \centering

  \includegraphics[width=\linewidth]{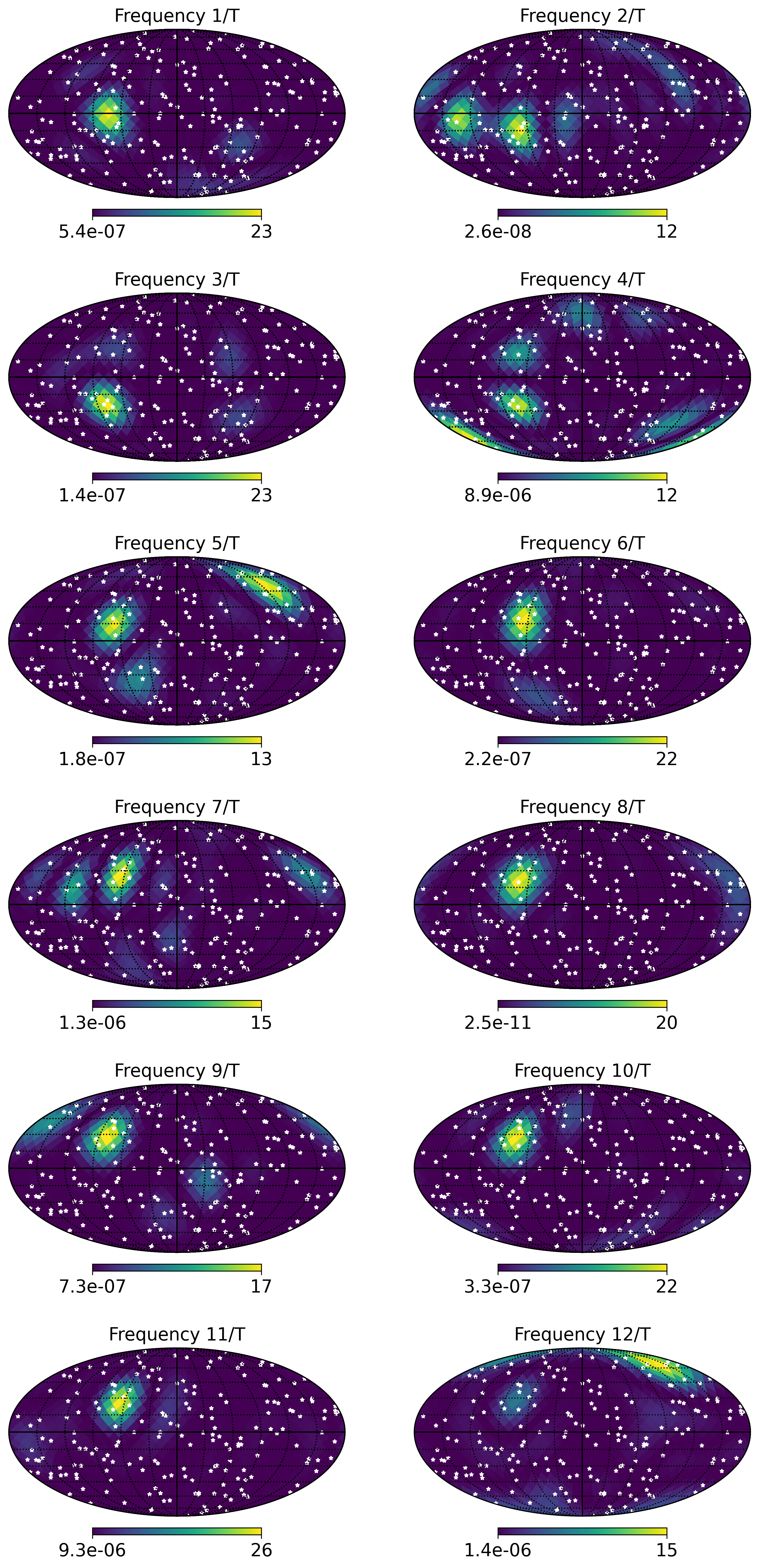}

  \caption{Sky maps of the GWB power distribution $P(\hat\Omega, f_j)$ recovered from simulated PTA datasets injecting one Pop 2 realization. White stars are $200$ pulsars randomly distributed in the sky. We show that in almost all frequencies, our pipeline identifies the correct location of the two brightest sources; the first one dominates the first 4 sky maps, and the second dominates up to frequency $11/T$.} \label{fig:sky_map_results}
\end{figure}

\subsection{Detection statistic for eccentricity}
\label{statistics}

Here, we quantify the detectability of eccentricity from sky map cross-correlation. To this end, we build a statistic to contrast the correlation matrix, $\bar{r}^{ij}$, of Eq.~\eqref{eq:equation} produced by a population of circular binaries (Pop 1, our null hypothesis) with that generated by the eccentric population Pop 2. We proceed as follows:
\begin{enumerate}
  \item We compute the matrix $\bar{r}^{ij}$ for 100 independent realizations of our PTA experiment injecting SMBHBs from the Pop 1 model, and we combine the 6600 cross-correlation values (considering that $\bar{r}^{ij}$ is a $12\times12$ matrix, there are $12\times11/2=66$ per realization) to construct a single correlation distribution $p(\bar{r})$ under the null hypothesis (i.e., quasi-circular binaries).
  \vspace{0.2cm}
  \item From this, we construct the null hypothesis cross-correlation cumulative distribution $\gamma(\bar{r})=\int_{-1}^{\bar{r}} p(r) dr$.
  \vspace{0.2cm}
  \item We use the following quantity as a proxy for estimating the probability of drawing 66 values $\bar{r}_k$ from the null hypothesis distribution
\begin{equation}
\label{eq:lik}
\Lambda = \prod_{k=1}^{66}[1-\gamma(\bar{r}_k)].
\end{equation}

  \item We calculate the probability distribution of $\Lambda$ in the null hypothesis $p(\Lambda_{\rm circ})$ by drawing 8 million times 66 correlation values from the null distribution $p(\bar{r})$ and we define the significance at which the null hypothesis can be rejected given a measured likelihood $\Lambda$ (i.e. $p$-value) as
\begin{equation}
\label{eq:pval}
S = \int_{-\infty}^{\Lambda} p(\Lambda_{\rm circ}).
\end{equation}
  
  \item Finally, we calculate the correlation matrix $\bar{r}^{ij}$ for 100 PTA experiments featuring Pop 2 in the injection and, for each of them, compute the probability $\Lambda_{\rm ecc}$ of extracting the 66 reconstructed $\bar{r}^{ij}$ correlation values from the null correlation distribution according to Eq.~\eqref{eq:lik}.
\end{enumerate}
If the correlation matrices of Pop 2 are significantly different from those derived under the null hypothesis (Pop 1), the values obtained of $\Lambda_{\rm ecc}$ will fall in the tail of the distribution $p(\Lambda_{\rm circ})$, and the null hypothesis can be rejected with high significance. We note that by computing a single $p(\bar{r})$ in step 1, we implicitly assume that all individual $\bar{r}^{ij}$ follow the same distribution in the null hypothesis. We verified that this is the case by testing each $p(\bar{r}^{ij})$ distribution against the overall $p(\bar{r})$ distribution with a Kolmogorov-Smirnov test, and we found no significant outliers.

\paragraph{Triplets.}

The statistic defined in the previous section is based only  on the values of the $\bar{r}^{ij}$ correlation coefficients, regardless of their location in the correlation matrix. However, the latter contains information about the nature of the signal, since eccentric binaries are expected to be more strongly correlated in adjacent frequency bins. This is due to the smooth shape of the function $g(n,e)$, which features an $e$-dependent peak at some $n$. 
In a first attempt to account for the expected larger correlation in adjacent frequency bins, we constructed a coefficient $C(i,j,j+1)$ that captures the correlation of the sky map at frequency $f_i$ with two sky maps at frequencies $f_j$, $f_{j+1}$ as
\begin{equation}
\label{eq:lik2}
C(i,j,j+1) = \bar{r}^{ij} + \bar{r}^{i,j+1} + \bar{r}^{j,j+1}.
\end{equation}
Using 12 frequency bins, we can construct 100 of these sky map correlation coefficient "triplets" instead of 66 $\bar{r}^{ij}$. To test whether this triplet correlation can better identify eccentricity, we constructed a triplets detection statistic following steps 1 to 5 of the previous section but starting from $C(i,j,j+1)$ instead of $\bar{r}^{ij}$.

\subsection{Results}
\label{pta}

In Figure \ref{fig:corr_pta} we show correlation matrices obtained from the PFOS 
presented in Section \ref{sec:freq} for the same SMBHB populations shown in Figure \ref{fig:corr_pta_theo}. For Pop 1, we get what we expected: diagonal values equal to one and off-diagonal elements close to zero. Looking at Pop 2, we still recover the correlations in the off-diagonal elements, even though we notice that the values are lower than the theoretical ones shown in Figure \ref{fig:corr_pta_theo} (see Section \ref{discussion} for more details). Despite this, we still recognize that two eccentric bright sources dominate the GWB, producing correlations in different sky maps. Those are evident in Figure \ref{fig:sky_map_results}, where we show that we can reconstruct the sky localization of these sources in almost all frequency bins, although with larger uncertainties and additional hot spots compared to Figure \ref{fig:sky_maps} (right panels).

\begin{figure*}

  \centering

  \begin{subfigure}[b]{0.5\linewidth}
    \centering
    \includegraphics[width=\linewidth]{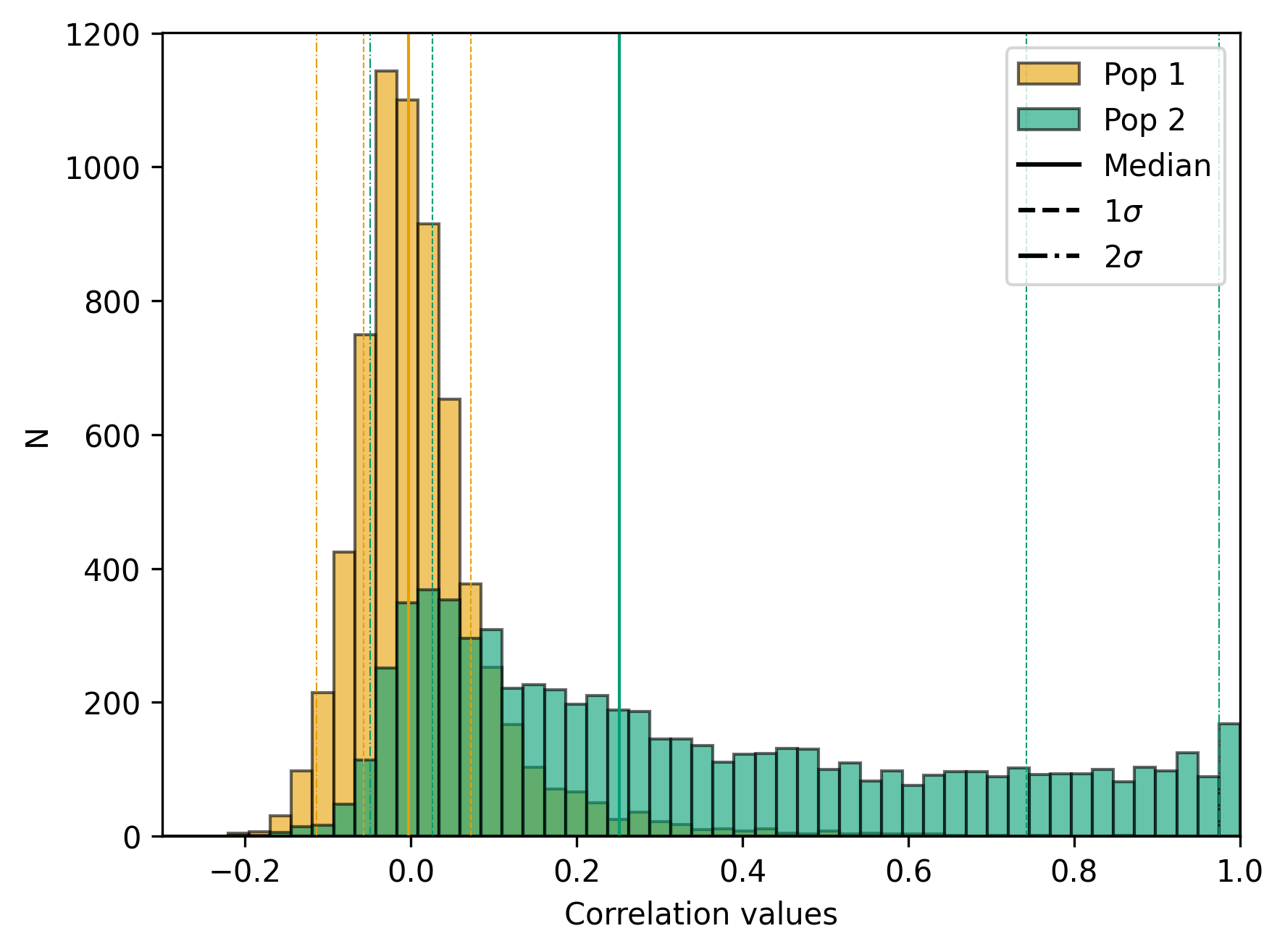}

  \end{subfigure}\hfill
  \begin{subfigure}[b]{0.5\linewidth}
    \centering
    \includegraphics[width=\linewidth]{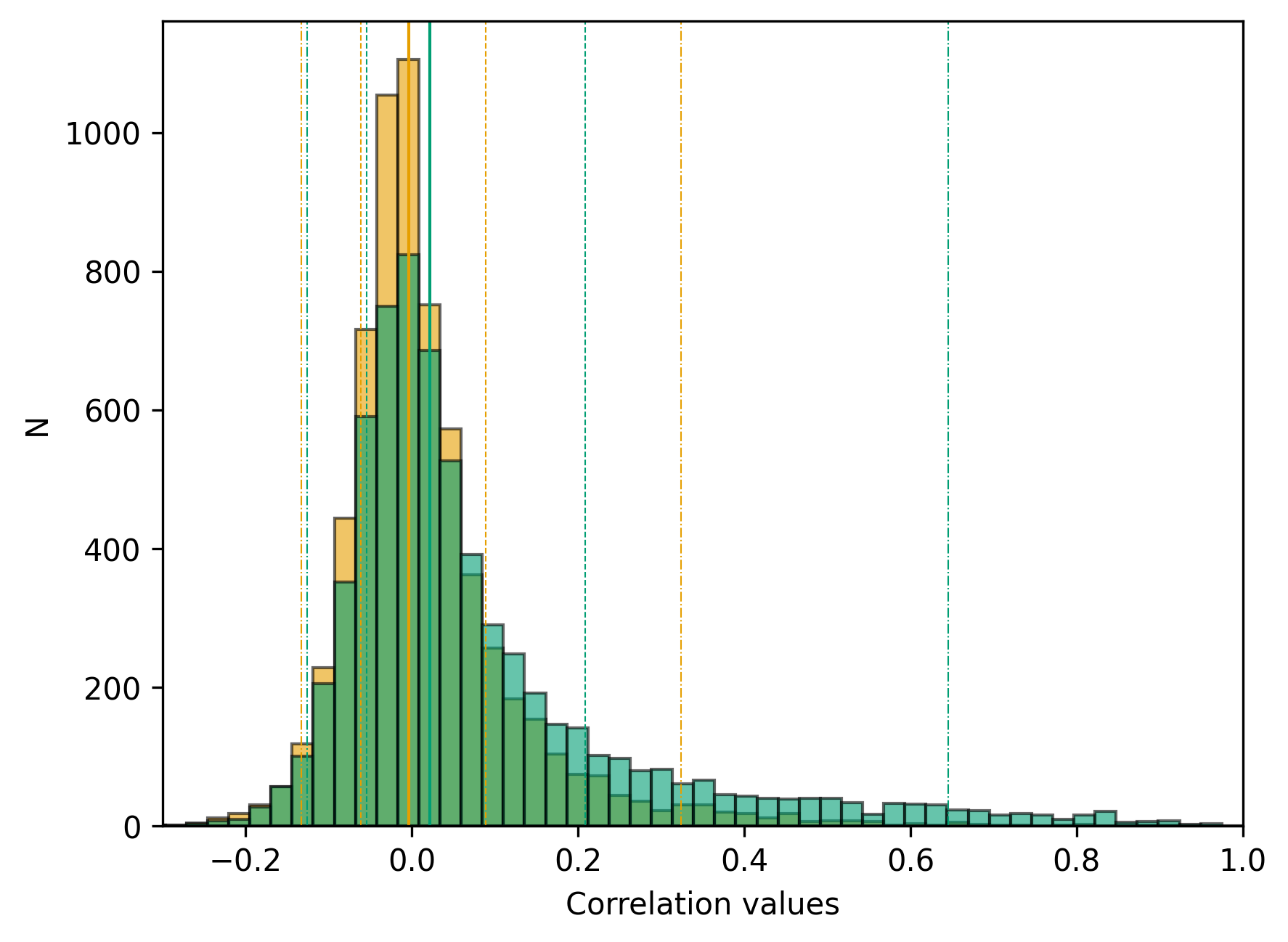}
  \end{subfigure}
  \caption{Correlation distributions of Pop 1 and Pop 2 realizations recovered from known SMBHB (left plot), and from simulated PTA datasets (right plot). \label{fig:correlations}}
\end{figure*}

The poorer quality of the reconstructed sky maps results in less distinctive correlation values for the circular and eccentric SMBHB populations, as shown in Figure \ref{fig:correlations}. While for Pop 1 there is virtually no difference in the distribution $p(\bar{r}^{ij})$ whether it is calculated from the theoretical sky maps of $h_c^2$ or it is reconstructed from a realistic PTA (orange histograms), this is not the case for Pop 2 (green histograms). The $p(\bar{r}^{ij})$ reconstructed from the PTA, while presenting a significant tail at positive correlation values, is an identical null hypothesis distribution. 

\begin{figure*}

  \centering

  \begin{subfigure}[b]{0.5\linewidth}
    \centering
    \includegraphics[width=\linewidth]{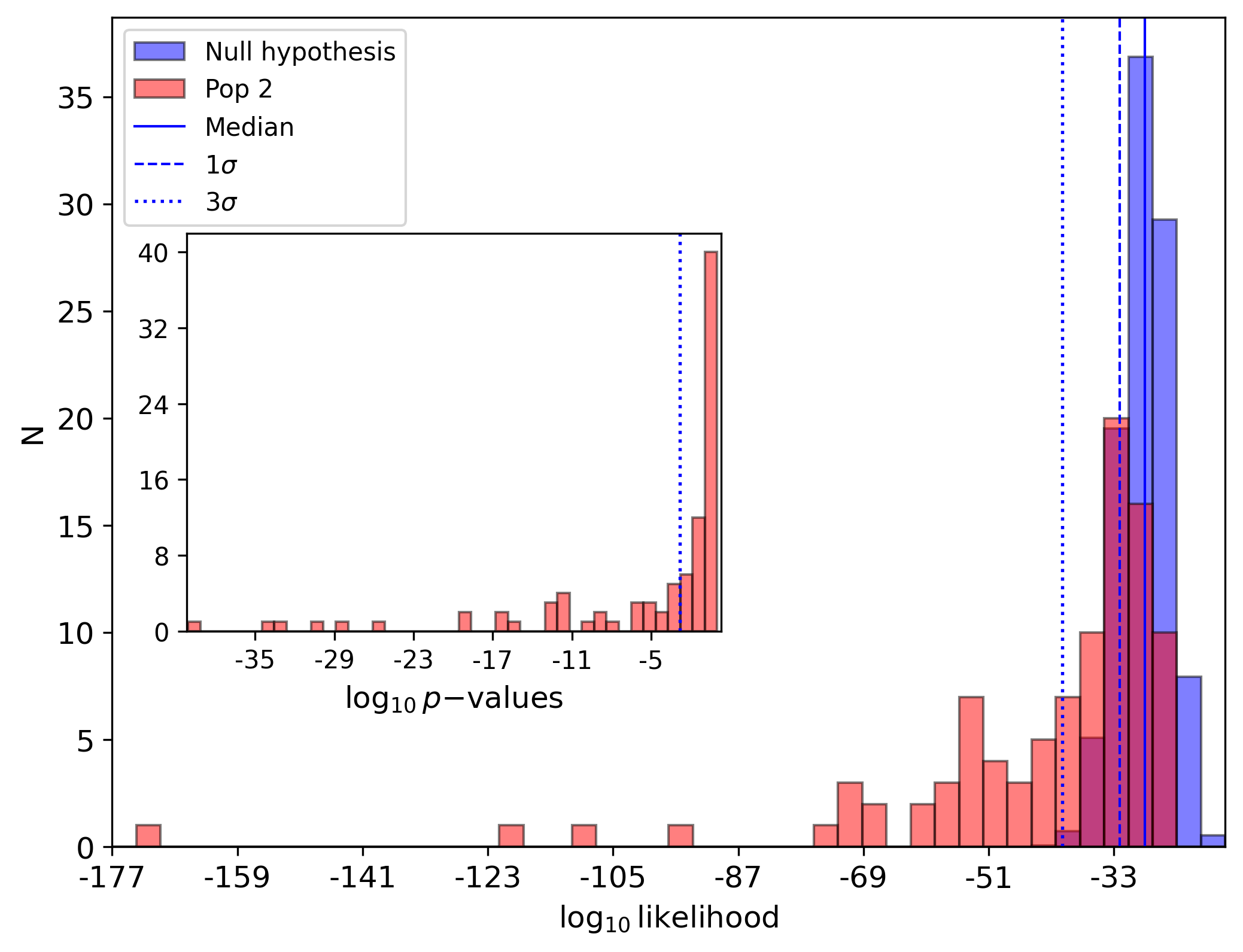}

  \end{subfigure}\hfill
  \begin{subfigure}[b]{0.5\linewidth}
    \centering
    \includegraphics[width=\linewidth]{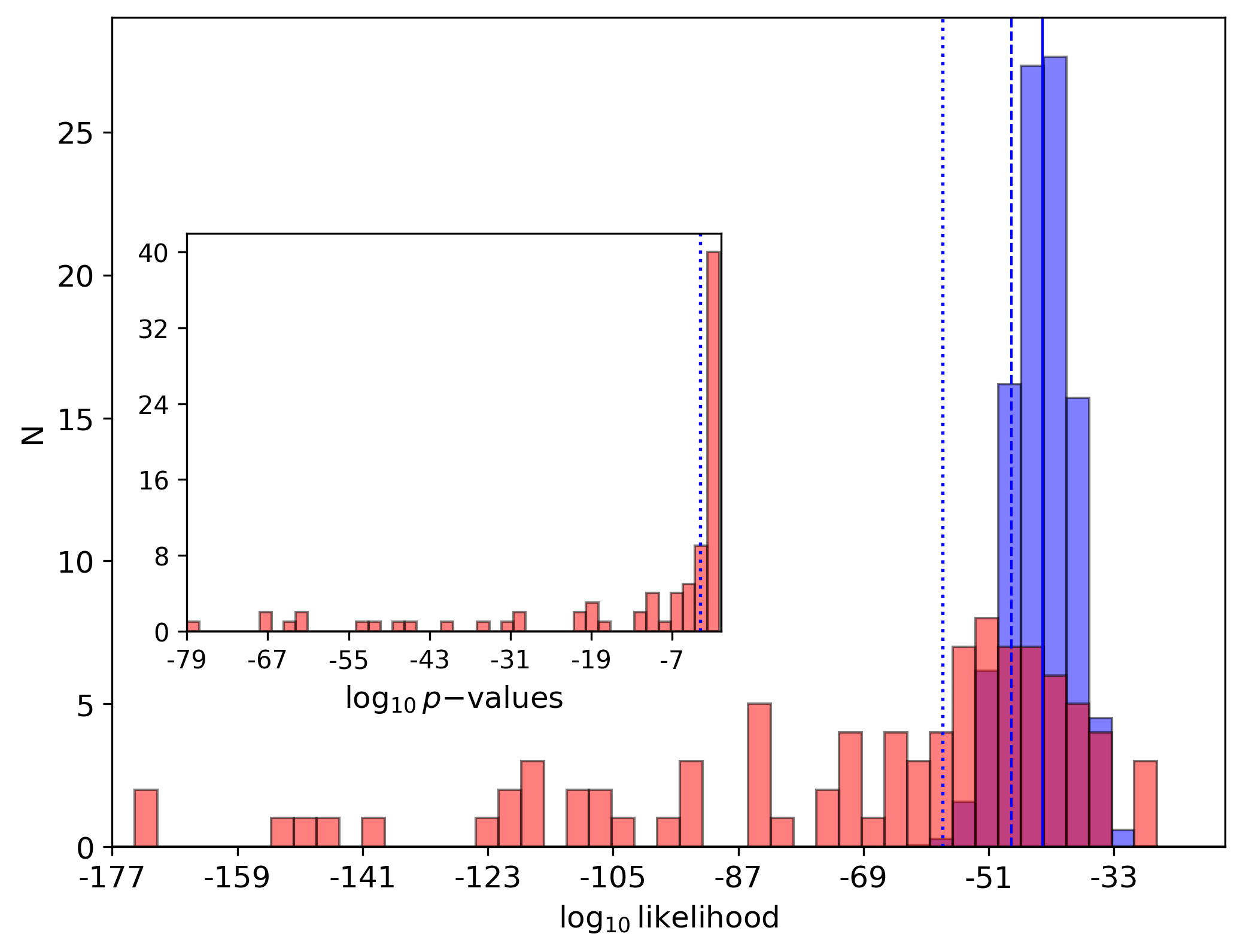}

  \end{subfigure}
  \caption{Likelihood distributions of Pop 2 realizations compared to the null hypothesis using pair (left plot) and triplet correlations (right plot). Note that 3 and 6 realizations of Pop 2 (in left and right plots, respectively) are not shown because their likelihood values are equal to zero, which implies strong correlation. The $p$-value distributions are also shown in the inset plots. There, 3 and 13 realizations of Pop 2 which have $p$-values equal to zero do not appear in the left and right plots, respectively.\label{fig:likelihood}}
\end{figure*}

To quantify the detectability of eccentricity, we then computed the correlation pairs and triplets statistics described in the previous section, testing 100 realizations of PTAs featuring a Pop 2 injection. The results are shown in Figure \ref{fig:likelihood}, where we compare the likelihood of these 100 eccentric models against the null hypothesis. For the correlation pair statistic (left panel) we observe that the $\Lambda_{\rm ecc}$ distribution is much more skewed toward a lower value but we get a $p$-value $S<1.4\times10^{-3}$ (i.e., a 3$\sigma$ rejection of the null hypothesis) only in $37\%$ of cases. The same threshold is exceeded in 52\% of the cases for triplet statistics (right panel), which indeed, by including the correlation of multiple adjacent bins, performs better.

\section{Discussion and conclusions \label{discussion}}

We developed a method to detect a correlated GW sky power distribution across the frequency spectrum by cross-correlating sky maps at different frequencies. This is particularly interesting because such correlations arise if the GWB present in the data is generated by a population of eccentric binaries that emit power across a broad frequency spectrum. Despite the limitations encountered analyzing ideal SKA-like PTA datasets (detailed below), our method can be used to distinguish a GWB produced by a quasi-circular SMBHBs population affected by environmental coupling from one generated by an eccentric population of SMBHBs featuring similar spectral attenuation at low frequencies.
We demonstrated the key concept by constructing the cross-correlation matrix of theoretical sky maps of the GW power only and then tested its validity on realistic (although highly idealized) PTA datasets. 

Although the method performs flawlessly on theoretical maps, it is less effective (although still valuable) on simulated data. This is because the reconstructed sky maps are much less precise, leading to correlation matrices that are less informative than the theoretical ones, as highlighted in Figure \ref{fig:correlations}.

There are multiple reasons that contribute to this lower performance. 
\begin{enumerate}

  \item The PTA observable is an incoherent superposition of GWs with different initial phases in the time domain. When multiple GWs of comparable amplitude contribute to overall power, their inherent incoherency is likely to eliminate correlations, resulting in lower values of $\bar{r}^{ij}$. Although this is certainly a limiting factor, it does not seem to be a no-go for real-life applications of this technique. In fact, we found that our correlation statistics perform much better when the number of pulsars in the array increases from 100 to 200. 
  \vspace{0.1cm}  
  \item An additional limiting factor is given by the size and properties of the adopted PTAs. Intrinsic noise, complicated and broad response functions, and sparse distribution of the pulsars all contribute to the degradation of the reconstruction of the sky distribution of the GW power. 
  \vspace{0.1cm}    
  \item 
  Another real-life complication that requires further investigation is the leakage of GW power observed in the same cases between consecutive frequency bins. In fact, the Fourier basis that describes the GWB signal is finite and has a resolution limit of $1/T$. Each SMBHB will contribute to the PSD estimate of the GWB at the discrete frequency $n/T$, which is closer to the frequency at which they are actually emitting GWs. However, when a GW source emits at a frequency that is in between two consecutive elements of the defined Fourier basis ($n/T$), its GW power contribution may "leak" into the neighboring bin. In other words, the GW source will contribute to the GW-power estimate at both Fourier components and to both sky maps, inducing some level of correlation between them, even if the studied binary is circular. \\

  \vspace{0.1cm}    
  \item Lastly, to construct the sky maps, we used the frequentist PFOS outside of its strict validity domain. In fact, in the presence of eccentric binaries, the GWB is non-stationary and the PFOS (as most PTA analysis tools) is not strictly applicable since the Fourier domain covariance matrix of the GWB is not diagonal. Although we showed that the reconstructed sky maps are reasonable (cf. Figure \ref{fig:sky_map_results}), more sophisticated reconstruction techniques need to be developed.
\end{enumerate}

We also caution that the SMBHB populations employed in this initial investigation were simplistic, featuring binaries that all shared the same eccentricity at formation, $e_0=0.01$ for Pop 1 and $e_0=0.9$ for Pop 2. Since anisotropies in the GWB power distribution are generally dominated by a few loud sources, the detection of correlation mostly constrains the properties of those loud systems, which might not be representative of the overall population. In an extreme (likely unrealistic) example, strong correlation might be detected even if a single loud eccentric source dominates the GW power emitted by a population of otherwise circular binaries. We defer tests involving more complex SMBHB populations to future work.

With anisotropy studies like this, we can constrain the sky position of multiple single sources, as well as the strain $h_c(f)$ given the computed power at different frequencies. Thus, this method is complementary to Bayesian and/or frequentist continuous gravitational waves (CGW) searches that use a circular and eccentric pipeline \cite[e.g.,][and references therein]{B_csy_2022,Ferranti_2025}. Moreover, through anisotropy studies, we can put constraints on the prior range of CGW searches, reducing the parameter space and the computation time, increasing the possibility of constraining other CGW parameters. Future projects will involve multiple CGW searches and the comparison of the results with anisotropy studies.

To conclude, we developed and tested this new analysis using a simulated PTA dataset, enabling the recovery of the expected correlation using eccentric and quasi-circular populations applying the PFOS tools. This method also allows us to recover the sky localization of the bright sources that dominate the GWB. Work is currently in progress to improve the pipeline and make it applicable to real PTA datasets.

\begin{acknowledgements}
We thank the B-Massive group at Milano-Bicocca University for useful discussions and comments. B.E.M, A.S, G.M.S, D.I.V, and A.C acknowledge the financial support provided under the European Union’s H2020 ERC Consolidator Grant “Binary Massive Black Hole Astrophysics” (B Massive, Grant Agreement: 818691) and the financial support provided under the European Union Advanced Grant “PINGU” (Grant Agreement: 101142079). AC acknowledges financial support from the European Research Council (ERC) starting grant ’GIGA’ (grant agreement number: 101116134). LS would like to acknowledge the support of the European Space Agency through ESA’s postdoctoral Research Fellowship programme.
\end{acknowledgements}

\bibliographystyle{aa} 
\bibliography{references}
\end{document}